\documentclass[sn-mathphys,Numbered]{sn-jnl}% Math and Physical Sciences Reference Style
\usepackage{graphicx}%
\usepackage{multirow}%
\usepackage{amsmath,amssymb,amsfonts}%
\usepackage{amsthm}%
\usepackage{mathrsfs}%
\usepackage[title]{appendix}%
\usepackage{xcolor}%
\usepackage{textcomp}%
\usepackage{manyfoot}%
\usepackage{booktabs}%
\usepackage{algorithm}%
\usepackage{algorithmicx}%
\usepackage{algpseudocode}%
\usepackage{listings}%
\usepackage[font=small,labelfont=bf]{caption}
\usepackage{subcaption}

\begin{document}

\title[Article Title]{Supervised and Unsupervised Deep Learning Approaches for EEG Seizure Prediction}

%%\dgr{MSc, PhD}}\email{iauthor@gmail.com}
%%=============================================================%%

\author[1,2]{\fnm{Zakary} \sur{Georgis-Yap}}\email{ztcgy11@gmail.com}
\equalcont{These authors contributed equally to this work.}

\author[1,2]{\fnm{Milos} \sur{R. Popovic}}\email{milos.popovic@uhn.ca}

\author*[1,2]{\fnm{Shehroz} \sur{S. Khan}}\email{shehroz.khan@uhn.ca}
\equalcont{These authors contributed equally to this work.}

\affil*[1]{\orgdiv{KITE}, \orgname{University Health Network}, \orgaddress{\street{550, University Avenue}, \city{Toronto}, \postcode{M5G 2A2}, \state{Ontario}, \country{Canada}}}

\affil[2]{\orgdiv{Institute of Biomedical Engineering}, \orgname{University of Toronto}, \orgaddress{\street{64 College St.}, \city{Toronto}, \postcode{M5S 3G9}, \state{Ontario}, \country{Canada}}}


\abstract{Epilepsy affects more than $50$ million people worldwide, making it one of the world’s most prevalent neurological diseases. The main symptom of epilepsy is seizures, which occur abruptly and can cause serious injury or death. The ability to predict the occurrence of an epileptic seizure could alleviate many risks and stresses people with epilepsy face. 
%Most of the previous work is focused at seizure detection, we pivot our focus to seizure prediction problem. 
We formulate the problem of detecting preictal (or pre-seizure) with reference to normal EEG as a precursor to incoming seizure. To this end, we developed several supervised deep learning approaches model to identify preictal EEG from normal EEG. We further develop novel unsupervised deep learning approaches to train the models on only  normal EEG, and detecting pre-seizure EEG as an anomalous event. These deep learning models were trained and evaluated on two large EEG seizure datasets in a person-specific manner. We found that both supervised and unsupervised approaches are feasible; however, their performance varies depending on the patient, approach and architecture. This new line of research has the potential to develop therapeutic interventions and save human lives.}




\maketitle

\section{Introduction}
Epilepsy is one of the most prevalent neurological disorders in the world, affecting approximately $1\%$ of the world's population \cite{Moh18, Beg18, Lit02}. Epilepsy is characterized by spontaneously occurring seizures, which %cause patients considerable stress. Epileptic seizures can also result in
could lead to bodily injuries, %(such as head injuries), 
fractures,  burns \cite{Ngu09}, and death in many cases %. Epileptic seizures can also lead to death, with $16\%$ of people with epilepsy dying during an epileptic seizure 
{\cite{Rid11}. %Therapeutic options to manage seizures exist. However, these options are ineffective for approximately $25\%$ of people with epilepsy \cite{Lit02}. Beyond the dangers of seizures, people with untreatable epilepsy must live with the uncertainty and stress that come with not knowing when their next seizure will occur. 
People with epilepsy are mostly concerned with
%reported that found that their number one reported negative impact was 
the fear of incoming seizures \cite{Fis00}. Therefore, there is a dire need to reduce the unpredictability of seizures to reduce the risk of injuries and improve
%for people with epilepsy so that the risk of injury is minimized and patients can have improved 
their quality of life.

%\subsection{Background}
Electroencephalography (EEG) is normally used to analyze brain activity pertaining to seizures \cite{Kuh18}. %It is also used by epileptologists to identify seizures, diagnose epilepsy, and treat patients \cite{Noa09}. %There are two main types of EEG: scalp EEG (sEEG) and intracranial EEG (iEEG) \cite{Noa09}. In sEEG, electrodes are placed on the surface of the scalp, while in iEEG, electrodes are inside the cranium \cite{Noa09}. iEEG can either be in the form of electrodes placed on the surface of the brain or electrodes implanted into the brain matter \cite{Noa09}.
Brain activity in people with epilepsy can be separated into four states: regular brain activity (interictal), brain activity before the seizure (preictal), brain activity during the seizure (ictal), and brain activity immediately after a seizure (postictal). The preictal state can contain 
%is particularly important to seizure prediction as it is widely agreed to be when there are 
observable physiological changes prior to the onset of a seizure \cite{Sta11} that can be used to predict an incoming seizure. 
The capability to predict an epileptic seizure could alleviate the risks patients face \cite{Bri16}; it %Injuries resulting from epileptic seizures are often due to them occurring at inopportune moments such as driving, cooking, or being at some sort of elevation \cite{Ngu09}. Being alerted to an incoming seizure 
would give patients the time to get help and %to a safe space or to find someone to accompany them, 
greatly reduce the risk of injury. %A further extension would be the development of a closed-loop system, where pre-emptive therapy is applied to a patient when a seizure is predicted \cite{Nag151}.
However, the biggest challenge is designing seizure prediction approaches is that there is no universally agreed upon preictal period length (PPL). Bandarabadi et al. \cite{Ban15} investigated the optimal PPL for seizure prediction using statistical analysis and found that the optimal PPL varies for each patient and for seizure within each patient \cite{Ban15}. %This makes it difficult for most prediction approaches as they must make a fixed assumption on the PPL.

Most of the work in this area is around seizure detection \cite{Gia14}, which involves detecting a seizure after its occurrence. Although this is important, contemporary work must aim to predict seizures before their onset, as it can save patients' lives and improve their quality of life. Our main hypothesis is that the correct detection of preictal state against normal brain activity (through supervised or unsupervised approaches) can be a strong indicator of an incoming epileptic seizure.
In the supervised setting, a binary classifier can be trained between interictal and preictal periods. Whereas, in the unsupervised setting, a classifier can be trained on only normal EEG (interictal) and preictal state can be identified as an anomaly. 
%explored various deep learning-based seizure prediction methods using EEG. The first approach was a supervised approach where we classified the preictal and interictal periods, with correctly classifying a preictal state as being an indicator of incoming seizure \cite{georgispreictal}. 
Our main contributions are:% as under:
\begin{itemize}
    \item Presented supervised and new unsupervised deep learning approaches to predict epileptic seizures.% before their occurrence.
    \item Experimentally determined the PPL and window size, as against heuristics or domain knowledge.
    \item Performed leave-one-seizure out cross validation for better generalization of results.
    \item All the experiments were performed in a patient-specific manner to avoid data leakages, overestimation of results and emphasis on individualized outcomes.
    %\item We experimented with two SWEC-ETHZ iEEG and the CHB-MIT sEEG datasets.
\end{itemize}
%We experimented both the approaches the SWEC-ETHZ iEEG dataset and the CHB-MIT sEEG dataset, in a patient-specific manner. %All of our implementations were patient-specific. 
%We experimented with the benefits of optimizing the window size and PPL of the model. We used a leave-one-seizure-out partitioning method to ensure our models were evaluated fairly. We were able to achieve good performance on the majority of the patients. We then moved to an unsupervised approach to alleviate the dependency on preictal data. We trained models solely on the the interictal data and anything anomalous was treated as preictal. Anomaly detection was done by attempting to reconstruct the input using autoencoders. 
Our results showed that the unsupervised approaches were able to obtain comparable results to supervised seizure prediction in many patients. However, across all implementations there was not one clear best-performing model. % and average results were not statistically significant enough to make any conclusions. Despite this, we demonstrated that unsupervised anomaly detection seizure prediction is a potentially viable approach that should be explored further. 
This paper is an extension of our preliminary work \cite{georgispreictal} that introduced supervised convolution neural network (CNN) on SWEC-ETHZ dataset \cite{Bur19}. In this paper, we present two new supervised approaches: CNN-Long Short Term Memory (LSTM) and Temporal Convolution Network (TCN), and three new unsupervised approaches (CNN, CNN-LSTM, TCN autoencoders). We developed new seizure prediction baselines for the SWEC-ETHZ dataset \cite{Bur19} and included a new CHB-MIT dataset \cite{Sho09}. 

%Therefore, we can formulate seizure prediction as a classification problem between the interictal state and the preictal state to indicate an incoming seizure \cite{Zha16, Shi17, georgispreictal}.

\section{Related Work}
% \subsection{EEG datasets}
% To develop seizure predictors, it is important to have access to large amounts of epilepsy data \cite{Kuh18}. Long-term, continuous EEG datasets that are available to the public have emerged over the last decade to supplement this need \cite{Kuh18}. A list of commonly used public human EEG datasets is shown in Table \ref{table:datasets}.

% \begin{table*}[t]
% \centering
% \begin{tabular}{ |c|c|c|c|c|c| } 
%  \hline
%  \textbf{dataset} & \textbf{Hours} & \textbf{Patients} & \textbf{Seizures} & \textbf{Type of EEG} & \textbf{Free?} \\ \hline
%  EPILEPSIAE   & ~45,000 & 275 & ~2,500 & Scalp and Intracranial & No \\ \hline
%  AES  & ~230    & 2   & 46     & Intracranial  & Yes \\ \hline
%  CHB-MIT      & ~900    & 23  & ~170   & Scalp & Yes \\ \hline
%  SWEC-ETHZ    & ~2500   & 18  & 116    & Intracranial  & Yes \\ \hline
%  Melbourne Competition & ~30,000 & 3   & ~1,100 & Intracranial  & Yes \\ \hline
% \end{tabular}
% \caption{Summary of commonly used public human EEG seizure datasets.}
% \label{table:datasets}
% \end{table*}

%\subsection{Machine Learning Seizure Prediction}
%Machine learning has become a commonly used technique to learn patterns from data and apply those learnings to new, unseen data in the same distribution \cite{ElA20, Kot07}. 
Seizure prediction using supervised machine learning %has emerged as the primary method 
has been used to distinguish the interictal and preictal states \cite{Ach18}. %Supervised machine learning is a subset of machine learning where data samples have class labels and a model learns to classify data into one of the possible classes \cite{Kot07}.
Typical supervised machine learning seizure prediction approaches involve signal pre-processing, extracting and selecting features, followed a classifier \cite{Ach18}. Common signal processing techniques include high-pass, low-pass, or band-pass filtering, as well as artifact removal techniques \cite{Ach18}. Feature extraction is typically done by a bio-signals or epilepsy expert examining a patient’s EEG and deciding appropriate features  for separating the preictal and interictal states \cite{Ach18}. These features are often patient-specific and include statistical, non-linear, frequency domain, and time-frequency domain features \cite{Ach18,natu2022review}. Common classifier choices include support vector machines (SVM), k-nearest neighbour and random forest \cite{Ach18}.

Machine learning approaches may have limitations in terms of extracting handcrafted features, which could be suboptimal and time-consuming. 
%A limitation of supervised machine learning for seizure prediction is that it often relies on handcrafted features to distinguish seizure states. Although it has been shown to be effective, the use of handcrafted features relies on an expert to analyze a patient's EEG, which is very time-consuming. Also, these feature extraction methods may not be transferable to new patients. This raises the problem of scalability, as it is not feasible to have an expert analyze each patient's data for every seizure prediction implementation. There is a need to instead rely on algorithms that can use the same or similar features for all patients or that can extract features without an expert.
%\subsection{Deep Learning Seizure Prediction}
Deep learning approaches can overcome some of these challenges %is a subset of machine learning that uses multiple non-linear neural network layers 
by able to learn features from data with little to no pre-processing, generate high-level representations of data, and learn complex functions \cite{LeC15}. }{However, deep learning approaches need vast amount of data and computing resources to train and deploy the models, which can be time-consuming and costly. Also, it requires a careful tuning of the hyperparameters to avoid overfitting and under fitting.%Deep learning can be applied effectively to bio-signals, which are known to be complex and dynamic \cite{Gan18}.
%Many recent EEG seizure prediction methods use deep learning approaches \cite{Ach18}. 
An overview of preictal-interictal classification seizure prediction methods (on human subjects) using deep learning is shown in Table \ref{table:lit_review}. %All of these works are done on human EEG.

\begin{table*}[t]
\centering
\begin{tabular}{|c|c|c|c|c|}
\hline
\textbf{Citation} & \textbf{Window Size} & \textbf{PPL} & \textbf{Pre-processing} & \textbf{DL Architecture} \\
      & (Seconds)   & (Minutes)    &        &  \\ \hline
~\cite{Tei14}  &  5   & 10,20,30,40  & Various fixed features  & MLP      \\ \hline
~\cite{Fei17}  &  10  & Unknown      & Fractional FT  & MLP      \\ \hline
~\cite{Mir09}  & 300 & 50  & Handcrafted    & CNN      \\ \hline
~\cite{Kha18}  &  1   & 10  & CWT    & CNN      \\ \hline
~\cite{Tru18}  & 30  & 60  & STFT   & CNN      \\ \hline
~\cite{Ebe18}  & 15  & 60  & None   & CNN      \\ \hline
~\cite{Zha20}  & 5   & 30  & Common spatial pattern  & CNN      \\ \hline
~\cite{YuL20}  &  30  & 60  & Fast FT & Multi-view CNN    \\ \hline
~\cite{Dis21}  &  10  & 60  & MFCC   & CNN      \\ \hline
~\cite{Jan21}  &  1,2,3,8     & 10  & None   & CNN      \\ \hline
~\cite{georgispreictal} &  5,10,15,30,60 & 30,60,120 & STFT & CNN \\ \hline
~\cite{XuY20}  &  20  & 30,60        & None   & CNN      \\ \hline
~\cite{Tsi18}  & 5   & 15  & Handcrafted    & LSTM     \\ \hline
~\cite{Moh18}  & 10  & 30  & STFT   & CNN + LSTM        \\ \hline
~\cite{Abd18}  & 4   & 60  & None   & CNN-AE+BiLSTM     \\ \hline
~\cite{Dao19}  &  5   & 60  & None   & MLP,CNN,CNN+LSTM  \\ \hline
~\cite{Wei19}  & 10  & 30  & Image conversion        & CNN+LSTM \\ \hline
~\cite{Usm20}  &  30  & 60  & STFT   & CNN+SVM  \\ \hline
~\cite{Hus20}  &  30  & 30  & None   & 1DCNN+GRU\\ \hline
~\cite{Pra21}  &  Unknown     & 60  & Image conversion        & 3DCNN    \\ \hline
~\cite{Ozc19}  &  4   & 30  & Various fixed features  & 3DCNN    \\ \hline
~\cite{Tru19}  & 28  & 30  & STFT   & Convolutional GAN \\ \hline
\end{tabular}
\vspace{0.2cm}
\caption{Overview of deep learning EEG seizure prediction methods.}
\label{table:lit_review}
\end{table*}

Many reviewed deep learning methods performed some type of pre-processing the EEG data before passing it on to the classifier, typically through filtering \cite{Moh18, Kha18}, artifact removal \cite{Pra21}, or time-frequency analysis \cite{Tru18, Kha18}. %These pre-processing methods are patient and dataset agnostic, meaning they do not need to be manually altered for each patient.
Common deep learning architectures used for seizure prediction include CNN \cite{Moh18, Tru18}, LSTM network \cite{Tsi18, Wei19}, and feed-forward multilayer perceptron (MLP) \cite{Tei14}. We observed that majority of the studies use CNNs, LSTMS and/or their combinations to benefit from learning spatial and temporal features. %which are known to be effective on image data, despite LSTMs being known for being effective for time-series data. One-dimensional EEG signals are converted into a two-dimensional ``image'', using a time-frequency transform or by stacking raw inputs from all channels together.
The window size (fixed duration of data to analyze) and PPL were kept fixed in most of the studies, and they varied even when working on the same dataset and patients. This is an issue in building classifiers to predict seizures because the optimal PPL varies across patients (as concluded by Bandarabadi et al. \cite{Ban15}). %This implies that any method using preictal-interictal classification for seizure prediction should invariably experiment on finding optimal PPL. %The majority of studies use a fixed PPL and provide no empirical rationale for their choice. 
Only four of the studies reported experimenting with the PPL \cite{Tei14, Tsi18, Kha18, Pra21}, while others did not present any rationale for their choices. %Another interesting finding is that 
Some of the studies (e.g., \cite{Sho09}) also found different PPLs sizes, showing that the optimal PPL varies depending on the method’s implementation. These studies show that it is better to determine the PPL empirically at a patient-specific level, rather than using a generic or pre-determined average over a population. %They also show that new preictal-interictal classification methods cannot simply reference previous studies' parameter choices because optimal PPLs may vary on a per-implementation basis. 

We extend the existing supervised methods by obtaining PPL and window size using a leave-one-seizure-out (LOSO) evaluation and introduced a new supervised TCN classifier for this task. %supervised classification 
%There is no known work on using unsupervised deep learning for seizure prediction using EEG. For the first time, 
}{Unsupervised deep learning approach has been used by Daoud et al. \cite{Dao19}; in conjunction with classification approach to predict epileptic seizures. They trained a deep autoencoder on unlabelled EEG segments (balanced data of preictal and interictal segments) and replace the trained autoencoder in their classification pipeline. Their intent to use autoencoder is to leverage the transfer learning abilities of the model by allowing the training process to have a good start point instead of random initialization of the parameters, which reduces the training time drastically. We introduced three different autoencoder models in a total unsupervised manner and separate from the classification approaches and studied their performance for this problem.

\section{Supervised Seizure Prediction}
Preictal-interictal classification for seizure prediction is performed with three different architectures: convolutional neural networks (CNN) (used in our previous work \cite{georgispreictal}), and two new architectures, i.e., CNN-LSTM), and TCN. We briefly discuss them below.

%\subsection{Model Description}
\subsection{CNN}
The CNN model takes in EEG samples that have been time-frequency transformed using a STFT \cite{Tru18} (see Section \ref{sec:stft}). This helps the model in extracting time and frequency features and puts the data into a suitable format for $2$D convolutions \cite{Tru18}. The CNN architecture takes advantage of spatial information in data to learn relevant features. %and typically requires less parameters than feed forward neural networks \cite{LeC95}.
Each sample was converted into a $2$D matrix \(F \times T\) using a STFT, where \(F\) was the number of sample frequencies used and \(T\) was the number of segment times used. The matrix was then resized to a \(128 \times 128\) ``image'' using bilinear interpolation so that image sizes were consistent regardless of the window size. The time-frequency transform was done independently for each channel, resulting in each sample being of dimensions \(C \times 128 \times 128\), where $C$ is the total number of channels.
The samples were then passed to the CNN model, which is made up of three convolutional blocks (see Figures \ref{fig:conv_block} and \ref{fig:s_cnn}), followed by three fully connected layers with ReLU activation functions. Table \ref{table:cnn_params} shows the model hyperparameters used for the CNN.

\begin{figure}[!htbp]
    \centering
    \begin{subfigure}{0.4\textwidth}(a)
    \centering
        \includegraphics[scale=0.7]{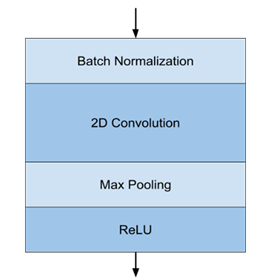}  \caption{}%figure}{Convolutional block.}
        \label{fig:conv_block}
    \end{subfigure}
    \hfill
    \begin{subfigure}{0.4\textwidth}(b)
    \centering
        \includegraphics[scale=0.7]{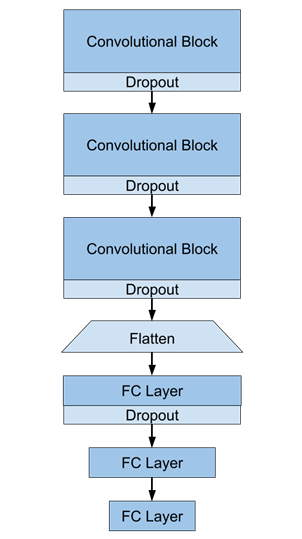} \\
        \caption{}%{CNN architecture with convolution blocks and fully connected (FC) layers.}
        \label{fig:s_cnn}
    \end{subfigure}
    \caption{(a) Convolutional block, (b) CNN architecture with convolution blocks and fully connected (FC) layers}
    \label{fig:cnn}
\end{figure}
%    \hfill

\begin{table*}[!htbp]
%    \begin{minipage}{0.3\textwidth}
    \centering
        \begin{tabular}{|p{2.1
        cm}|c|} \hline
   \textbf{CNN kernel size}       & 5\\ \hline
   \textbf{CNN filter sizes}      & 8, 16, 32 \\ \hline
   \textbf{Fully connected sizes} & 128, 64   \\ \hline
   \textbf{Dropout}      & 0.5       \\ \hline
        \end{tabular}
        \captionof{table}{CNN model hyperparameters.}
        \label{table:cnn_params}
%    \end{minipage}
\end{table*}

\subsection{CNN-LSTM}
The CNN-LSTM architecture takes advantage of both the spatial feature extraction of the CNN along with the LSTM's propensity to work well with temporal data. The CNN-LSTM model takes in STFT images similar to the CNN model. The input is a consecutive series of images as one sample. The input sequence is divided into smaller sub-sequences, which are %. Then each sub-sequence is 
independently time-frequency transformed and resized into  \(64 \times 64\) images, leading to dimensions \(C \times n \times 64 \times 64\), where \(n\) is the number of sub-sequences in a sample and is equal to the sequence length divided by the sub-sequence length.
Each sub-window is passed into a CNN model with two convolutional blocks that outputs a feature vector. Then, each feature vector %from each sub-window 
is concatenated into a sequence and passed into a $2$-layer LSTM, whose outputs are passed to a fully connected layer that outputs the final scores. An overview of the CNN-LSTM architecture and hyperparameters are shown in Figure \ref{fig:s_cnnlstm} and Table \ref{table:cnnlstm_params}.% shows the model hyperparameters used for the CNN-LSTM.
We did experiments with LSTM on raw EEG data; however, the results were not satisfactory and not discussed in the paper. Converting the EEG to STFT highlighted the frequencies and enabled CNN to learn spatial information, while the LSTM was able to model the temporal dependencies between there. The RNNs were not used because LSTM supports longer memory and better handling of vanishing gradient problem \cite{sherstinsky2020fundamentals}. On the flip side, LSTMs generally require more parameters to train, making it computationally expensive and memory intensive. We resolved this issue by training LSTM models on a high-performance computation GPU cluster (see Section \ref{sec:ExperimentalSetting}). 

\begin{figure}
    \centering
        \includegraphics[angle=270, scale=0.4]{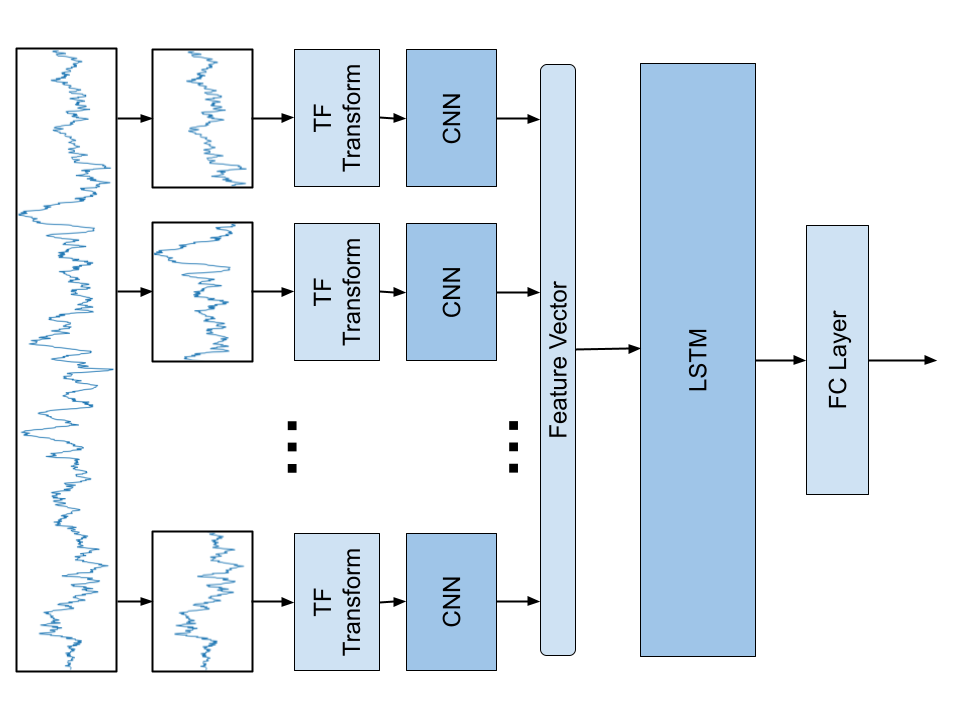} \\
        \captionof{figure}{CNN-LSTM architecture showing time-frequency (TF) transform, CNN layers, LSTM and fully connected (FC) layer.}
        \label{fig:s_cnnlstm}
\end{figure}

\subsection{TCN}
The TCN model takes in scaled sequences of size \(C \times S/4\), where \(S\) is the sequence length and the sequences were down-sampled by a factor of $4$. The TCN model \cite{Thi21} consisted of TCN blocks (see Figure \ref{fig:tcn}). Each TCN block is two consecutive sub-blocks that contain a causal $1$D convolution layer with a dilation, a weight normalization layer, a ReLU activation function, and a dropout layer \cite{Thi21}. The TCN blocks have skip connections, where the input to the block is added to the output \cite{Thi21}. The model contained $6$ TCN blocks with $32$ channels each, followed by a $1$D convolution layer, and a fully connected layer. The dilation factor of each block was }{\(2^{(n-1)}\), where \(n\) is the layer number. Figure \ref{fig:tcn} and Table \ref{table:tcn_params} shows the TCN architecture and hyperparameters.

\begin{figure}[!htbp]
    \centering
    \begin{subfigure}{0.4\textwidth}
        \includegraphics[scale=0.7]{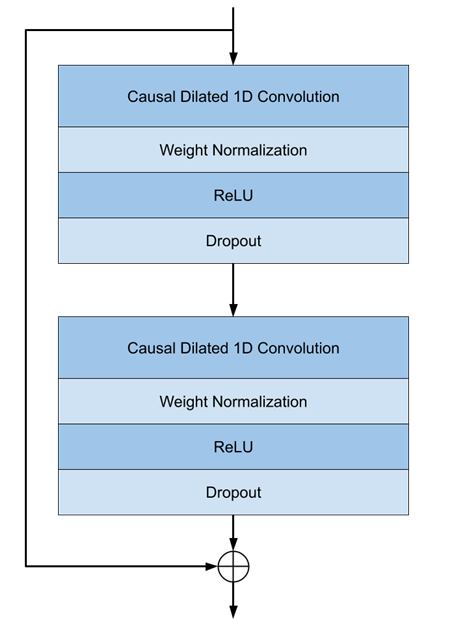} \\
        \caption{}
        %\label{fig:tcn_block}
    \end{subfigure}
    \hfill
    \begin{subfigure}{0.4\textwidth}
    \centering
        \includegraphics[scale=0.7]{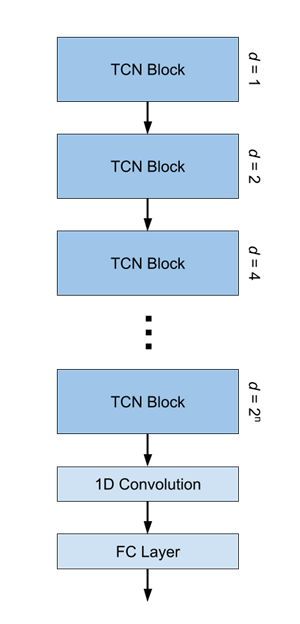} \\
        \caption{}
        %\label{fig:s_tcn}
    \end{subfigure}
    \caption{(a) TCN block description. TCN: temporal convolutional network. ReLU: rectified linear unit. (b) Supervised TCN architecture overview. TCN: temporal convolutional network. FC layer: fully connected layer.}
    \label{fig:tcn}
\end{figure}

\begin{table}[!htbp]    %\begin{table}
    %\centering
    \begin{subtable}{.45\linewidth}
        \begin{tabular}{ |c|c| } \hline
   \textbf{CNN kernel size} & 5     \\ \hline
   \textbf{CNN filter sizes}& 8, 16 \\ \hline
   \textbf{LSTM feature vector size} & 32    \\ \hline
   \textbf{LSTM hidden size}& 16    \\ \hline
   \textbf{Fully connected size}     & 96    \\ \hline
   \textbf{Dropout} & 0.5   \\ \hline
        \end{tabular}
        \caption{}
        \label{table:cnnlstm_params}
    \end{subtable}
    \qquad
    \begin{subtable}{.45\linewidth}
        \begin{tabular}{ |p{2.5cm}|p{1.5cm}| } \hline
   \textbf{TCN kernel size}       & 5     \\ \hline
   \textbf{TCN filter sizes}      & 32, 32, 32, 32, 32, 32 \\ \hline
   \textbf{Fully connected size}  & 64    \\ \hline
   \textbf{Dropout}      & 0.2   \\ \hline
        \end{tabular}
        \caption{}
        \label{table:tcn_params}
    \end{subtable}
    \caption{Hyperparameters of (a) CNN-LSTM, and (b) TCN models}
\end{table}

\section{Unsupervised Seizure Prediction}
The reliance on preictal data for supervised seizure prediction methods remains a challenge. Preictal data is typically scarce, and deep learning methods require a considerable amount of data from both classes to work well. Preictal-interictal classification methods cannot be used effectively on patients with little preictal data, and %even for patients with enough preictal data, 
class imbalance still remains an impending problem. An unsupervised approach (anomaly detection) to seizure prediction could remedy these problems. Anomaly detection for seizure prediction would require only interictal (and no preictal data) to train, making it easier to be more accessible to a larger population. Autoencoders (AEs) and its variants are apt to be used within this framework, with reconstruction error used as an anomaly score \cite{Denkovski2023}. To our knowledge, this is one of the first seizure prediction work that uses unsupervised deep learning approach for epileptic seizure prediction without utilizing preictal data. %based anomaly detection approach. %We have explored the viability of this approach.
%\subsection{Model Description}
% \subsubsection{Autoencoder}
% An autoencoder consists of an encoder that reduces the input to an embedding state and a decoder that attempts to reconstruct the input based on the embedding state \cite{Che18}. A general overview of the autoencoder architecture is shown in Figure \ref{fig:ae}. The architecture is forced to learn underlying patterns in the data to adequately reconstruct the input. Three different autoencoder architectures are used in this work.
%
% \begin{figure}[h]
%     \centering
%     \includegraphics[scale=0.25]{Figures/AE.png} \\
%     \caption{Basic autoencoder structure.}
%     \label{fig:ae}
% \end{figure}
%
%\subsubsection{CNN Autoencoder}
We implemented the following autoencoder approaches for this task. 
\begin{itemize}
    \item CNN autoencoder \cite{khan2021anomaly}.  Similar to the supervised CNN, it takes STFT images as input. The encoder is made up of three convolutional blocks followed by a fully connected layer which generates an embedding state of size $64$. The decoder is a mirrored version of the encoder (see %. An overview of the model is shown in 
    Figure \ref{fig:ae_cnn}).

    \item CNN-LSTM autoencoder \cite{nogas2018fall}. Similar to the supervised CNN-LSTM, the input sequence was divided into smaller sub-sequences and then an STFT was performed on each sub-sequence. The encoder consisted of an individual CNN encoder for each sub-sequence, followed by an LSTM that generated an embedding state of size $64$. The decoder has the reverse architecture to the encoder. %was the reverse: an LSTM followed by individual CNN decoders for each sub-sequence. An overview of the architecture is shown in 
    (see Figure \ref{fig:ae_cnnlstm}).

    \item TCN autoencoder \cite{abedi2022detecting}. It takes in raw scaled sequences, as is the case with the supervised TCN. The encoder was a TCN with three layers, each with $16$ channels followed by a $1$d convolution and a fully connected layer. The size of the embedding state was $64$. The decoder was an exact mirror of the encoder (see %. An overview of the model is shown in 
    Figure \ref{fig:ae_tcn}).
\end{itemize}

%The first architecture used for anomaly detection was the CNN autoencoder \cite{khan2021anomaly}. Similar to the supervised CNN, it takes in STFT images as input. The encoder is made up of three convolutional blocks followed by a fully connected layer which generates an embedding state of size $64$. The decoder is a mirrored version of the encoder. An overview of the model is shown in Figure \ref{fig:ae_cnn}.
%\subsubsection{CNN-LSTM Autoencoder}
%The second architecture used was the CNN-LSTM autoencoder. Similar to the supervised CNN-LSTM, the input sequence was divided into smaller sub-sequences and then an STFT was performed on each sub-sequence. The encoder consisted of an individual CNN encoder for each sub-sequence followed by an LSTM that generated an embedding state of size $64$. The decoder was the reverse: an LSTM followed by individual CNN decoders for each sub-sequence. An overview of the architecture is shown in Figure \ref{fig:ae_cnnlstm}.
%\subsubsection{TCN Autoencoder}
%The third architecture used was the TCN autoencoder. It takes in raw scaled sequences as is the case with the supervised TCN. The encoder was a TCN with three layers each with 16 channels followed by a one-dimensional convolution and a fully connected layer. The size of the embedding state was 64. The decoder was an exact mirror of the encoder. An overview of the model is shown in Figure \ref{fig:ae_tcn}.

\begin{figure}[!htbp]
    \centering
    \begin{subfigure}{0.5\textwidth}
        \centering
        \includegraphics[scale=0.22]{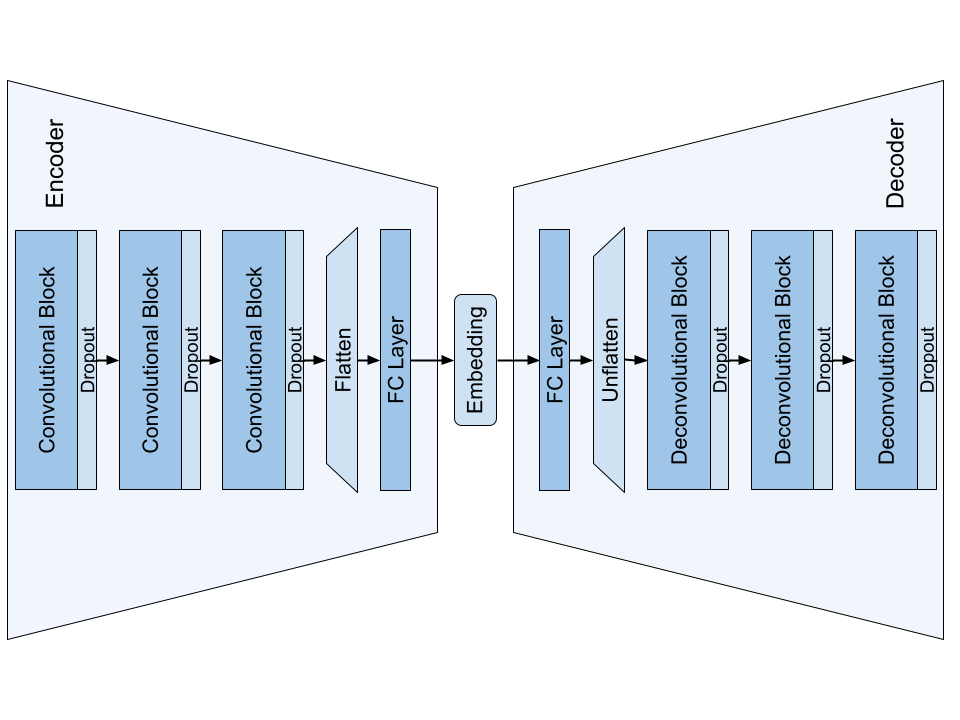}
        \caption{}
        \label{fig:ae_cnn}
    \end{subfigure}
    \begin{subfigure}{0.5\textwidth}
        \centering
        \includegraphics[scale=0.22]{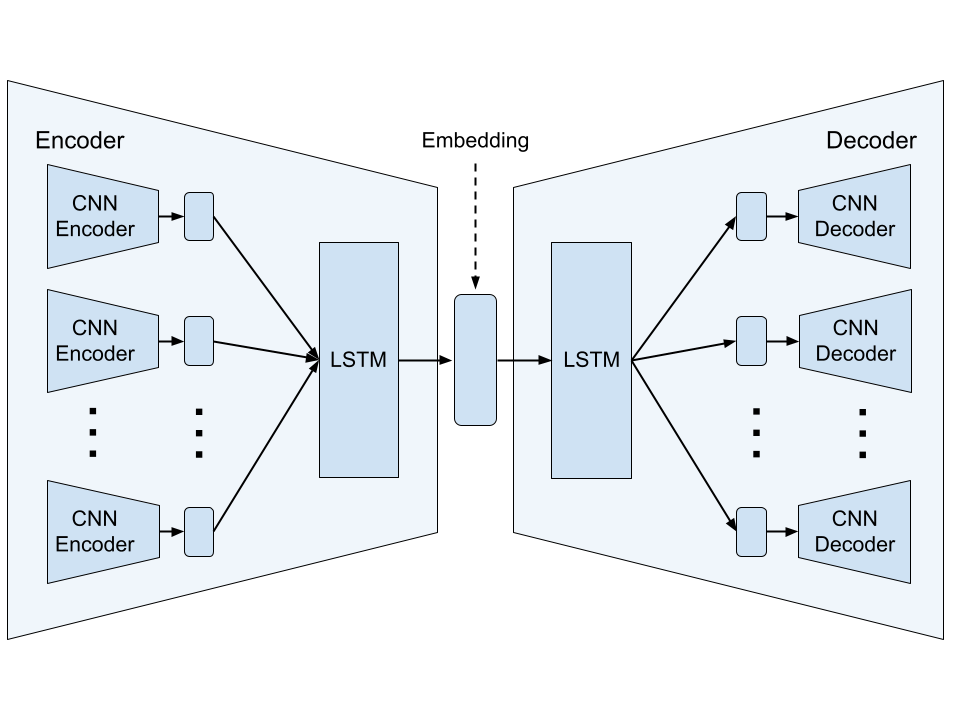}
        \caption{}
        \label{fig:ae_cnnlstm}
    \end{subfigure}
    \begin{subfigure}{0.5\textwidth}
        \centering
        \includegraphics[scale=0.22]{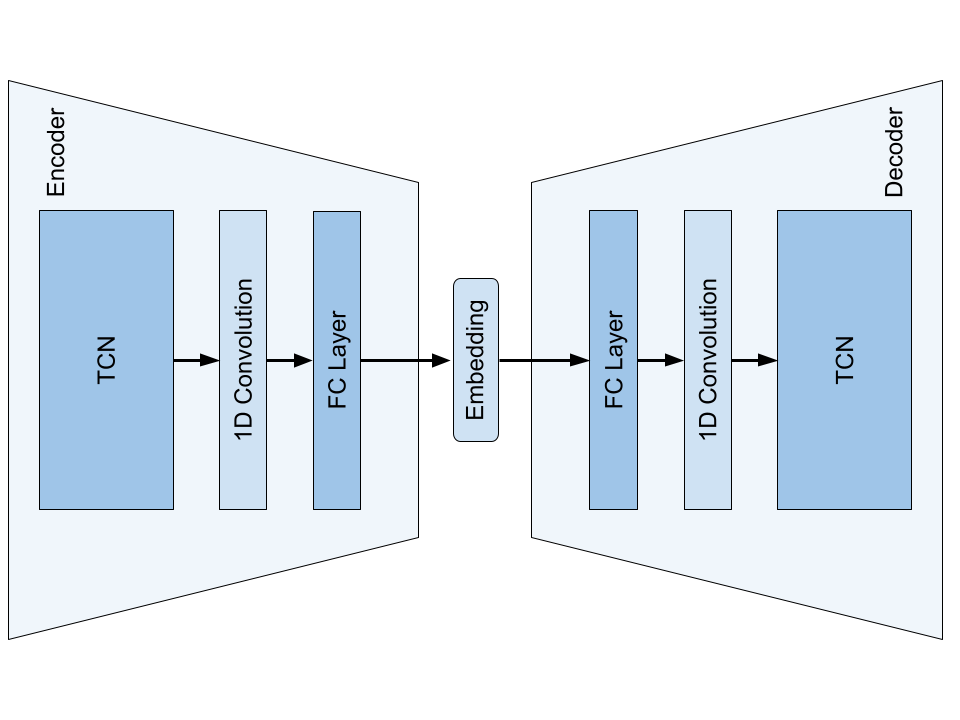}
        \caption{}
        \label{fig:ae_tcn}
    \end{subfigure}
\caption{Autoencoders comprising of (a) Convolution layers only, (b)CNN-LSTM, and (c) TCN}
\end{figure}

\section{Data Processing}
\subsection{Datasets}
We used two EEG Epilepsy seizure datasets, the Sleep-Wake Epilepsy Centre ETH Zurich (SWEC-ETHZ) dataset \cite{Bur19} and the Children's Hospital Boston Massachusetts Institute of Technology (CHB-MIT) dataset \cite{Sho09}. Both datasets are publicly available, easy to access, and contain human raw EEG recordings, where no seizure states have been pre-selected. This is important so we can define and experiment with different preictal and interictal regions. The SWEC-ETHZ dataset is an iEEG dataset containing over $2,500$ hours of recordings across $18$ patients with a sampling rate of either $512$Hz or $1024$Hz \cite{Bur19}. The CHB-MIT dataset contains scalp EEG recordings from $22$ patients sampled at $256$Hz with at least $22$ EEG electrodes \cite{Sho09}. }{Note that one patient had their recordings taken on two separate occasions $1.5$ years apart, and the two cases are treated as two separate patients for the rest of this paper \cite{Sho09}.

We define a ``lead seizure`` as any seizure that occurs at least $30$ minutes after a preceding seizure \cite{Tru18}. Only preictal periods from lead seizures were considered because of the lack of interictal and preictal data to train models. Patients that have less than three lead seizures were withheld from the experiments because at least three lead seizures were required to perform test partitioning combined with an internal leave-one-seizure-out (LOSO) cross-validation step (see Figure \ref{fig:partition}). Six out of the $18$ patients in the SWEC-ETHZ dataset were not considered for this work due to this condition. All patients in the CHB-MIT dataset had at least three lead seizures. A description of dataset attributes for all patients used from both the SWEC-ETHZ and CHB-MIT datasets is shown in Tables \ref{table:swec_overview} and \ref{table:chb_overview}, respectively.

\begin{table*}[!htp]
\centering
\begin{tabular}{ |c|c|c|c|c| } 
 \hline
 \textbf{Patient ID} & \textbf{Hours of data} & \textbf{Seizures} & \textbf{Lead seizures} & \textbf{Electrodes} \\ \hline
 ID03 & 158 & 4  & 4  & 64  \\ \hline
 ID04 & 41  & 14 & 14 & 32  \\ \hline
 ID05 & 110 & 4  & 4  & 128 \\ \hline
 ID06 & 146 & 8  & 8  & 32  \\ \hline
 ID07 & 69  & 4  & 4  & 75  \\ \hline
 ID08 & 144 & 4  & 4  & 61  \\ \hline
 ID09 & 41  & 23 & 14 & 48  \\ \hline
 ID10 & 42  & 17 & 15 & 32  \\ \hline
 ID12 & 191 & 9  & 9  & 56  \\ \hline
 ID13 & 104 & 7  & 7  & 64  \\ \hline
 ID16 & 177 & 5  & 5  & 34  \\ \hline
 ID18 & 205 & 5  & 5  & 42  \\ \hline
\end{tabular}
\caption{SWEC-ETHZ dataset patient description \cite{Bur19}}
\label{table:swec_overview}
\end{table*}

\begin{table*}[!htp]
\centering
\begin{tabular}{ |c|c|c|c|c| } 
 \hline
 \textbf{Patient ID} & \textbf{Hours of data} & \textbf{Seizures} & \textbf{Lead seizures} & \textbf{Electrodes} \\ \hline
 1  & 41  & 7  & 7  & 22 \\ \hline
 2  & 35  & 3  & 3  & 22 \\ \hline
 3  & 38  & 7  & 7  & 22 \\ \hline
 4  & 156 & 4  & 3  & 22 \\ \hline
 5  & 39  & 5  & 5  & 22 \\ \hline
 6  & 67  & 10 & 7  & 22 \\ \hline
 7  & 67  & 3  & 3  & 22 \\ \hline
 8  & 20  & 5  & 5  & 22 \\ \hline
 9  & 68  & 4  & 3  & 22 \\ \hline
 10 & 50  & 7  & 7  & 22 \\ \hline
 11 & 35  & 3  & 3  & 22 \\ \hline
 12 & 24  & 40 & 11 & 22 \\ \hline
 13 & 33  & 12 & 7  & 22 \\ \hline
 14 & 26  & 8  & 6  & 22 \\ \hline
 15 & 40  & 20 & 14 & 22 \\ \hline
 16 & 19  & 10 & 5  & 22 \\ \hline
 17 & 21  & 3  & 3  & 22 \\ \hline
 18 & 36  & 6  & 5  & 22 \\ \hline
 19 & 30  & 3  & 3  & 22 \\ \hline
 20 & 28  & 8  & 6  & 22 \\ \hline
 21 & 33  & 4  & 4  & 22 \\ \hline
 22 & 31  & 3  & 3  & 22 \\ \hline
 23 & 27  & 7  & 3  & 22 \\ \hline
\end{tabular}
\caption{CHB-MIT dataset patient description \cite{Sho09}}
\label{table:chb_overview}
\end{table*}

\subsection{Data Preprocessing}
The length and location of the preictal period is  defined by the PPL and the intervention time (IT). The IT is the time between the preictal state and the seizure onset. Interictal data is defined as any data that is not preictal, ictal, postictal, and is \(d\) distance away from the preictal state, as %. An example displaying these parameters is 
shown in Figure \ref{fig:params}.
The data was divided into samples of a fixed window size, % and each sample 
which were labelled as either interictal or preictal. We set \(d=0\) to evaluate the model’s ability to classify interictal and preictal samples in close temporal proximity to actual seizures. The IT was set to $0$, increasing it can be a future experiment after generating a baseline. In the SWEC-ETHZ dataset, interictal samples were randomly selected with a down-sampling factor of $8$ because interictal data were overly abundant, and the classes were significantly imbalanced (patients ID04, ID09, and ID10 used a down-sampling factor of $2$ instead because there was less interictal data). The number of preictal samples were artificially increased by using $50\%$ overlapping windows. % with an overlap size of half the length of each sample. 
The size of each sample was \(sf \times C\) where \(s\) was the window size, \(f\) was the sampling rate, and \(C\) was the number of EEG electrodes.

The dataset was partitioned into a training set and a testing set using LOSO partitioning. We used the last lead seizure's preictal data as the test set, while all other preictal data was part of the training set. As shown in Figure \ref{fig:partition}, LOSO partitioning is a better way to evaluate a model's ability to generalize to a new seizure's preictal data. Standard test partitioning where samples are randomly assigned to the training or test set may be an overestimation of the actual performance of the classifier. 

\begin{figure}[!htpb]
    \centering
    \begin{subfigure}{0.5\textwidth}
        \hspace*{-1cm}
        \centering
        \includegraphics[scale=0.6]{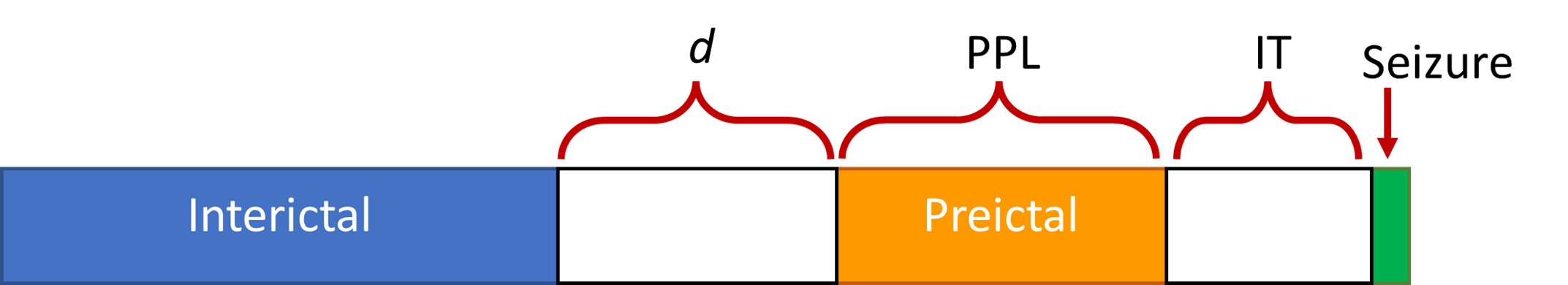}
        \caption{}
        \label{fig:params}
    \end{subfigure}
    \begin{subfigure}{0.5\textwidth}
        \hspace*{-4cm}
        \centering
        \includegraphics[scale=1]{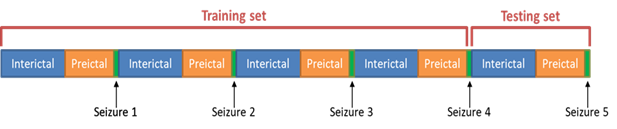}
        \caption{}
        \label{fig:partition}
    \end{subfigure}
    \caption{(a) Labelling of the preictal and interictal periods with parameters. (b) Simplified visualization of LOSO test partitioning by withholding the last seizure.}
\end{figure}

\subsection{Time-Frequency Transform}
\label{sec:stft}
We transformed the EEG data  %is often analyzed with the help of time-frequency analysis which involves transforming a signal 
from a time-series input into the time-frequency domain \cite{AlF14, Her14} %. A common time-frequency transform technique is the 
using short-time Fourier transform (STFT). It converts a one-dimensional time-series signal into a two-dimensional matrix of values with axes of time and frequency \cite{Her141}. The STFT %It does this by 
splits the signal into a series of smaller sequences and then performing Fourier transforms on each one individually, providing a way to see changes in the frequency domain at various points in time \cite{Fad12}. }{In this work, the sequence is split into segments of 128 samples before performing Fourier transforms.
In CNN based models used in the work, an STFT was used to pre-process the input before passing samples to the model. Other time-frequency analysis methods such as the continuous wavelet transform \cite{Kha18} and phase-amplitude coupling \cite{edakawa2016detection} were experimented with in our preliminary work but did not provide better results.

\begin{figure}[!htbp]
    \centering
    \includegraphics[scale=0.8]{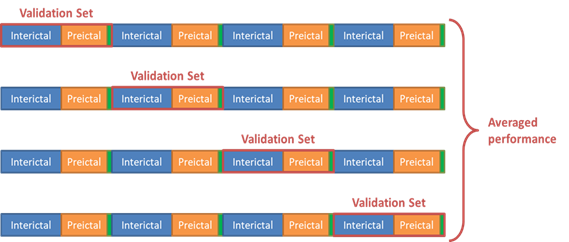} \\
    \caption{LOSO cross-validation example with four seizures. One seizure is used for validation, while the others are used for model training.}
    \label{fig:cv}
\end{figure}

\section{Experimental Setting and Results}
\label{sec:ExperimentalSetting}
%\subsection{Training and Validation}
A grid search was performed to find the optimal window size and PPL for each patient. We ran the model with varying window size ($5, 10, 15, 30, 60$ seconds) and PPL ($30, 60, 120$ minutes) values. We used an internal LOSO cross-validation to tune the parameters without looking at test data. This was done by dividing the training set into folds, where each fold was a different seizure's preictal and interictal data. One fold was the validation set, while the others were used for training. Each fold in the set was used as the validation set once, and the performance across all runs in a patient was averaged. An example of the cross-validation method used is shown in Figure \ref{fig:cv}. The  %metric we used to evaluate a model's performance was the 
area under the Receiver Operating Characteristic curve (AUC ROC) \cite{Hua05} was used as a performance metric for hyperparameter tuning. The test set was completely withheld from this process.
All the models were trained using an NVIDIA V100S-PCIe GPU with $32$ GB memory. A class-weighted (class weights vary per patient) cross-entropy loss function was used with the Adam optimizer and was trained for $100$ epochs with a batch size of $128$ and a learning rate of $0.0001$. %The data was in 32-bit floating point format. 
All implementations were done in the PyTorch framework \cite{Pas19}.
After the final parameters for a model were set, it was evaluated on the test set using the AUC ROC and precision-recall curve (AUC PR). AUC PR is more appropriate for imbalanced classification problems \cite{Bra16}. The reported performance metrics are AUC ROC and AUC PR. The calculation of false positive, false negative and derived metrics, such as accuracy, specificity, and sensitivity depend on the choice of the threshold applied to the reconstruction error or classification probabilities \cite{Dao19,Kha18}. ROC and PR analysis is more thorough since it does not depend on the threshold choice, and instead analyzes the specificities and sensitivities at all possible thresholds and provides an overall summary metric in terms of AUC ROC/AUC PR. For imbalanced datasets, AUC PR is a better metric as it takes care of precision and recall at all thresholds and does not inflate the results. Another important point to note is that in unsupervised deep learning models, there is no validation set; therefore, calculating an operating threshold is non-trivial.

\subsection{Supervised Prediction}
%\subsubsection{Hyperparameter Tuning Results}
Hyperparameter tuning results using the supervised CNN are shown in Tables \ref{table:results_swec_tuning} and \ref{table:results_chb_tuning} for the SWEC-ETHZ and CHB-MIT datasets, respectively. The window size and PPL obtained using cross-validations as well as AUC ROC vary considerably across different patients in both datasets. More than half of the patients in each dataset show AUC ROC values greater than $0.7$. %In some patients, validation AUC ROC was very high but then dropped when looking at the test AUC ROC. 
In the SWEC-ETHZ dataset, six of the patients had a test AUC ROC at least $0.1$ lower than their validation AUC ROC, while in the CHB-MIT dataset it was eight patients. This is consistent with Bandarabadi et al. \cite{Ban15} that the optimal preictal period for seizure prediction varies even on seizures within the same patient. The best way to account for this problem is to train and test on as many lead seizures' preictal data as possible.

\begin{table*}[!htbp]
\centering
\begin{tabular}{ |p{15mm}|p{15mm}|p{15mm}|p{17mm}|p{15mm}|p{15mm}| } 
 \hline
 \textbf{Patient ID} & \textbf{Window Size} & \textbf{PPL (seconds)} & \textbf{Validation AUC ROC} & \textbf{Test AUC ROC} & \textbf{Test AUC PR} \\ \hline
 ID03 & 30 & 1800 & 0.793 & \textbf{0.939} & 0.681 \\ \hline
 ID04 & 60 & 3600 & 0.708 & 0.509 & 0.610 \\ \hline
 ID05 & 60 & 7200 & 0.953 & \textbf{0.918} & 0.863 \\ \hline
 ID06 & 60 & 7200 & 0.704 & \textbf{0.948} & 0.842 \\ \hline
 ID07 & 30 & 7200 & 0.722 & 0.713 & 0.745 \\ \hline
 ID08 & 15 & 7200 & 0.722 & 0.454 & 0.452 \\ \hline
 ID09 & 30 & 7200 & 0.901 & \textbf{0.944} & 0.982 \\ \hline
 ID10 & 60 & 3600 & 0.807 & 0.574 & 0.790 \\ \hline
 ID12 & 60 & 1800 & 0.981 & \textbf{0.798} & 0.544 \\ \hline
 ID13 & 60 & 1800 & 0.721 & 0.499 & 0.394 \\ \hline
 ID16 & 10 & 1800 & 0.719 & 0.423 & 0.160 \\ \hline
 ID18 & 15 & 1800 & 0.832 & \textbf{0.850} & 0.401 \\ \hline
\end{tabular}
\caption{Validation and test results for preictal-interictal classification with optimized hyperparameters on the SWEC-ETHZ dataset.}
\label{table:results_swec_tuning}
\end{table*}

\begin{table*}[!htbp]
\centering
\begin{tabular}{ 
|p{15mm}|p{15mm}|p{15mm}|p{17mm}|p{15mm}|p{15mm}| } 
%|c|c|c|c|c|c| } 
 \hline
 \textbf{Patient ID} & \textbf{Window Size} & \textbf{PPL (seconds)} & \textbf{Validation AUC ROC} & \textbf{Test AUC ROC} & \textbf{Test AUC PR} \\ \hline
 1  & 60 & 1800 & 0.987 & \textbf{0.997} & 0.991 \\ \hline
 2  & 5  & 7200 & 0.718 & \textbf{0.982} & 0.869 \\ \hline
 3  & 10 & 1800 & 0.853 & \textbf{1.000} & 1.000 \\ \hline
 4  & 5  & 7200 & 0.376 & 0.649 & 0.037 \\ \hline
 5  & 5  & 3600 & 0.821 & \textbf{0.828} & 0.555 \\ \hline
 6  & 10 & 3600 & 0.616 & \textbf{0.858} & 0.796 \\ \hline
 7  & 5  & 7200 & 0.744 & 0.133 & 0.113 \\ \hline
 8  & 10 & 1800 & 0.999 & 0.379 & 0.378 \\ \hline
 9  & 10 & 7200 & 0.789 & \textbf{0.788} & 0.471 \\ \hline
 10 & 5  & 1800 & 0.732 & 0.686 & 0.185 \\ \hline
 11 & 30 & 3600 & 0.890 & \textbf{0.978} & 0.803 \\ \hline
 12 & 60 & 7200 & 0.917 & 0.549 & 0.974 \\ \hline
 13 & 5  & 3600 & 0.973 & \textbf{0.898} & 0.394 \\ \hline
 14 & 5  & 1800 & 0.817 & 0.139 & 0.151 \\ \hline
 15 & 30 & 7200 & 0.824 & 0.470 & 0.806 \\ \hline
 16 & 60 & 1800 & 0.686 & 0.688 & 0.294 \\ \hline
 17 & 5  & 7200 & 0.933 & 0.565 & 0.221 \\ \hline
 18 & 60 & 1800 & 0.750 & \textbf{0.820} & 0.100 \\ \hline
 19 & 30 & 7200 & 1.000 & \textbf{0.966} & 0.930 \\ \hline
 20 & 5  & 1800 & 0.983 & \textbf{0.898} & 0.497 \\ \hline
 21 & 5  & 7200 & 0.807 & 0.721 & 0.200 \\ \hline
 22 & 5  & 7200 & 0.770 & 0.397 & 0.059 \\ \hline
 23 & 30 & 1800 & 1.000 & 0.614 & 0.662 \\ \hline
\end{tabular}
\caption{Validation and test results for preictal-interictal classification with optimized hyperparameters on the CHB-MIT dataset.}
\label{table:results_chb_tuning}
\end{table*}

\subsubsection{Comparison with Fixed Parameters}
We implemented a preictal-interictal classification model with a fixed window size of $30$ seconds and PPL of $1$ hour to compare to our tuned hyperparameter model. The CNN model architecture is identical to the optimized parameter implementation. This was done to explore the benefits of optimizing hyperparameters for seizure prediction. 
Figures \ref{fig:results_swec_fixedopt} show the comparison of the two methods on the SWEC-ETHZ and CHB-MIT dataset. In general, for the SWEC-ETHZ dataset, the optimized hyperparameter implementation performed slightly better than the fixed parameter. In patient ID09, the optimized hyperparameter implementation performed much better than the fixed parameter implementation. For patient ID09, the hyperparameter tuning found a window size of $30$ seconds and a PPL of $2$ hours. It is likely that there was additional preictal information in the extra hour of data not used in the fixed parameter implementation.
%Figure \ref{fig:results_chb_fixedopt} shows the same comparison but for the 
For the CHB-MIT dataset, most patients had similar results for both the fixed and optimized hyperparameter implementations. There were a few patients (ID $5$, $16$, $17$, $18$) that had much better results with the optimized model. However, there were also patients (ID $9$, $22$, $23$) who performed better with a fixed hyperparameter implementation. For these patients, the last seizure's optimal hyperparameters were likely different from the optimal hyperparameters for the preceding seizures in the patient's dataset. Figures \ref{fig:results_swec_fixedopt_pr} and \ref{fig:results_chb_fixedopt_pr} show the comparison between the optimized and fixed hyperparameter implementations for the SWEC-ETHZ and CHB-MIT datasets respectively using AUC PR instead. It can be observed that the optimized implementation generally performs better on the SWEC-ETHZ dataset in both metrics, and that the difference is marginal in the CHB-MIT dataset. These experiments indicate that hyperparameter tuning can potentially improve the performance in comparison to fixed parameters.

\begin{figure}[!htbp]
    \centering
    \includegraphics[scale=0.5]{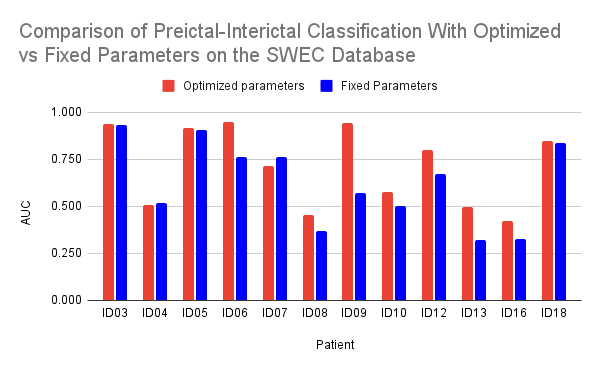} \\
    \caption{AUC ROC Comparison of CNN models using optimized hyperparameters vs fixed hyperparameters on the SWEC-ETHZ dataset.}
    \label{fig:results_swec_fixedopt}
\end{figure}

\begin{figure}[!htbp]
    \centering
    \includegraphics[scale=0.5]{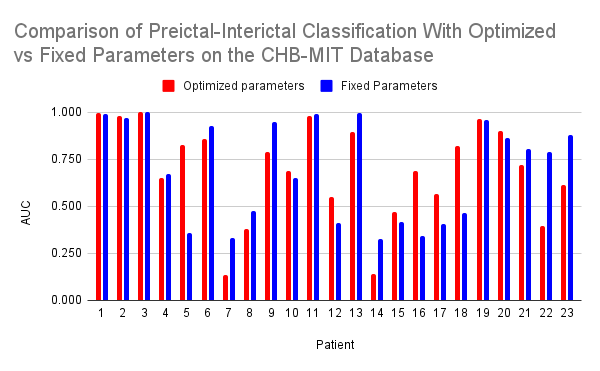} \\
    \caption{AUC ROC Comparison of CNN models using optimized hyperparameters vs fixed hyperparameters on the CHB-MIT dataset.}
    \label{fig:results_chb_fixedopt}
\end{figure}

%We additionally show AAUC PR) as an evaluation metric. AUC PR is recommended specifically for imbalanced data, which is common in seizure prediction problems. 
\begin{figure}[!htbp]
    \centering
    \includegraphics[scale=0.5]{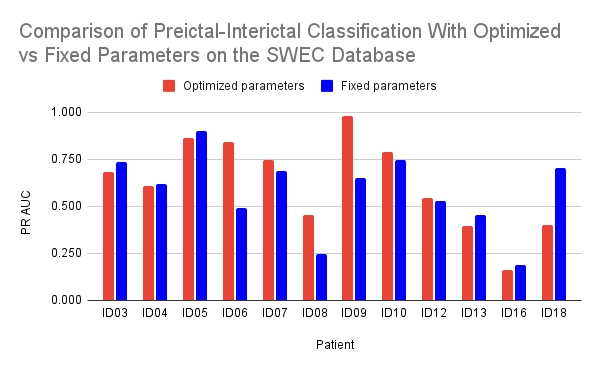} \\
    \caption{AUC PR Comparison of CNN models using optimized hyperparameters vs fixed hyperparameters on the SWEC-ETHZ dataset.}
    \label{fig:results_swec_fixedopt_pr}
\end{figure}

\begin{figure}[!htbp]
    \centering
    \includegraphics[scale=0.5]{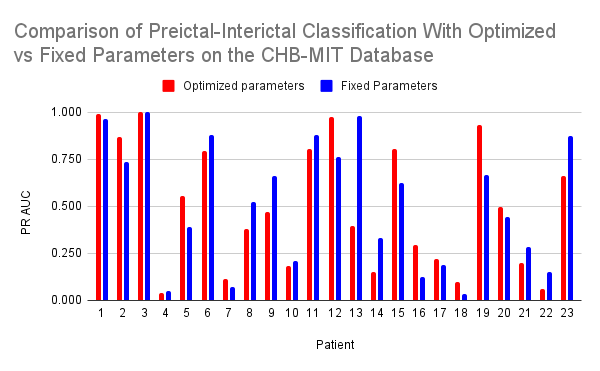} \\
    \caption{AUC PR Comparison of CNN models using optimized hyperparameters vs fixed hyperparameters on the CHB-MIT dataset.}
    \label{fig:results_chb_fixedopt_pr}
\end{figure}

\subsubsection{Comparison with Other Architectures}
%Implementations using other architectures have been developed to be compared with the CNN architecture. 
Using a fixed hyperparameter implementation, CNN-LSTM and TCN models were trained using a window size of $30$ seconds and a PPL of $1$ hour. A comparison of the AUC PR for all models is shown in Figures \ref{fig:results_swec_modelcomp_pr} and \ref{fig:results_chb_modelcomp_pr}. %The implementations vary in performance relative to each other depending on the patient.
%Looking at the AUC PR, we can observe which architecture performed best for a patient and the number of times a model type was the best performing for a patient.
In the SWEC-ETHZ dataset, the CNN, CNN-LSTM and TCN were the best performing model for $3$, $5$ and $4$ patients. %, the CNN-LSTM was the best performing model for $5$ patients, and the TCN was the best performing model for $4$ patients. 
The CNN and CNN-LSTM performed comparably well, and in each patient, the results were fairly similar. The TCN results were more variable, performing well on some patients, while the other models performed poorly. % on and vice versa. 
In the CHB-MIT dataset, the CNN, CNN-LSTM and TCN were the best performing model for $7$, $6$ and $10$ patients. %, the CNN-LSTM was the best performing model for $6$ patients, and the TCN was the best performing model for 10 patients. 
The TCN model performed much better in the CHB-MIT dataset compared to the SWEC-ETHZ dataset. Overall, the CHB-MIT results were very variable with AUC PR values, varying considerably even within the same patient.

\begin{figure}[!htbp]
    \centering
    \includegraphics[scale=0.5]{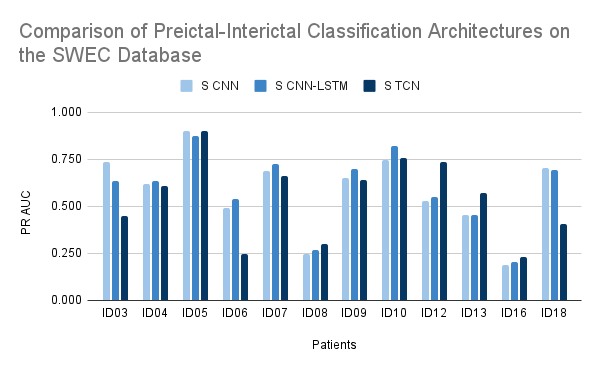} \\
    \caption{Comparison of CNN, CNN-LSTM, and TCN implementations for preictal-interictal classification on the SWEC-ETHZ dataset.}
    \label{fig:results_swec_modelcomp_pr}
\end{figure}

\begin{figure}[!htbp]
    \centering
    \includegraphics[scale=0.5]{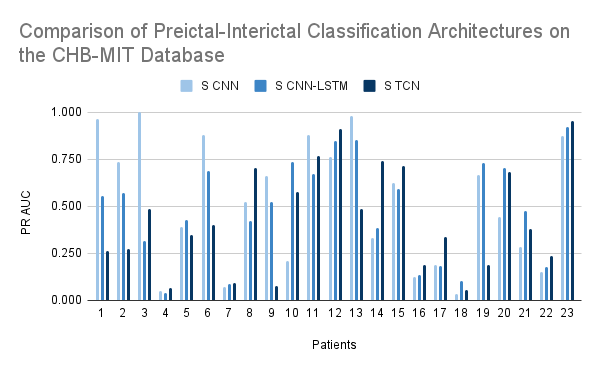} \\
    \caption{Comparison of CNN, CNN-LSTM, and TCN implementations for preictal-interictal classification on the CHB-MIT dataset.}
    \label{fig:results_chb_modelcomp_pr}
\end{figure}

\subsection{Unsupervised Prediction}
%\subsubsection{Training and Validation}
In the unsupervised approach, the training set only contained interictal data. For these experiments, the hyperparameters were fixed, with a window size of $30$ seconds and a PPL of $60$ minutes. %The models were trained using a NVIDIA V100S-PCIe GPU with 32GB memory and used the Adam optimizer with a mean-squared error loss function. 
The models were trained for $500$ epochs with a batch size of $128$ and a learning rate of $0.0005$.
After training, the models were evaluated on the test set that contained both interictal and preictal samples. Both AUC ROC and AUC PR were used to evaluate performance.

\subsubsection{Comparison of Architectures}
Figures \ref{fig:ucomp_swec_pr}, \ref{fig:ucomp_chb1_pr}, and  \ref{fig:ucomp_chb2_pr} show the anomaly detection based seizure prediction AUC PR results for the CNN, CNN-LSTM, and TCN AEs on the SWEC-ETHZ dataset and CHB-MIT dataset, respectively. We also show the supervised CNN with fixed hyperparameters for comparison. It can be observed that the performance varies significantly across different architectures and patients. For the SWEC-ETHZ dataset, the CNN AE performed the worst across most patients while the CNN-LSTM and TCN AEs performed relatively better, and even surpassed the supervised implementation in some patients. In the CHB-MIT dataset, the results vary even more, with no clear winner.

\begin{figure}[!htbp]
    \centering
    \includegraphics[scale=0.5]{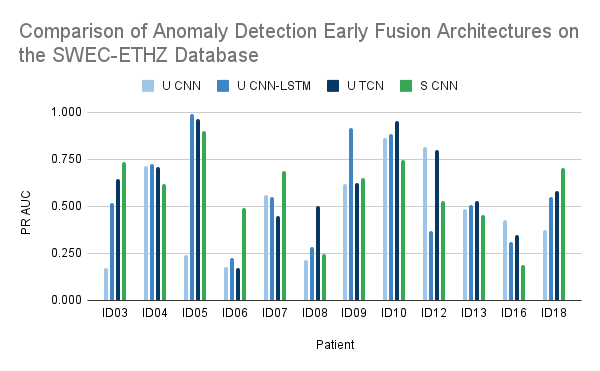} \\
    \caption{Comparison of unsupervised seizure prediction using different model architectures on the SWEC-ETHZ dataset. U: unsupervised, S: supervised.}
    \label{fig:ucomp_swec_pr}
\end{figure}

\begin{figure}[!htbp]
    \begin{subfigure}{0.5\textwidth}
        \centering
    \includegraphics[scale=0.5]{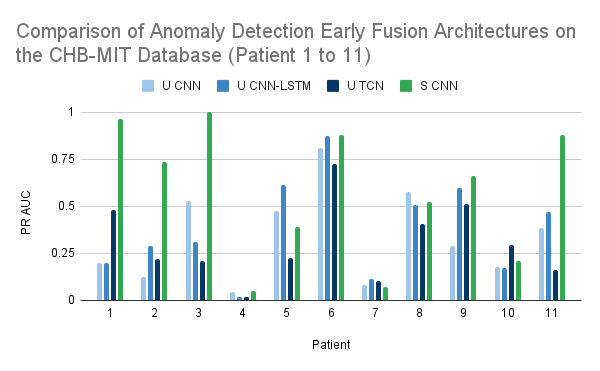} \\
    \caption{}
    \label{fig:ucomp_chb1_pr}
    \end{subfigure}
    \\
    \begin{subfigure}{0.5\textwidth}
    \centering
    \includegraphics[scale=0.5]{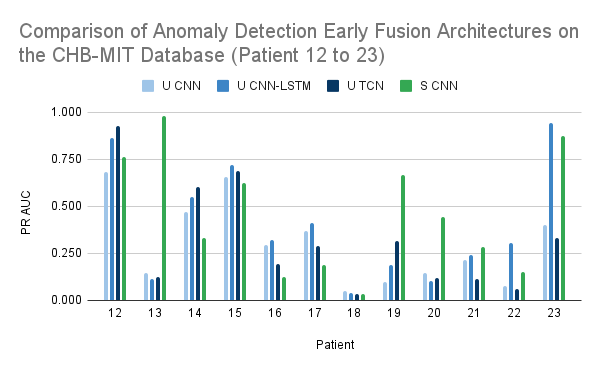} \\
    \caption{}
    \label{fig:ucomp_chb2_pr}
    \end{subfigure}
    \caption{Comparison of unsupervised seizure prediction using different model architectures on the CHB-MIT dataset (a) patients 1 to 11, (b) patients 12 to 23. U: unsupervised, S: supervised.}
\end{figure}

\subsection{Best Implementations}
Tables \ref{table:bestimpl_swec} and \ref{table:bestimpl_chb} show the best-performing implementation (from all experiments with supervised and unsupervised approaches) for each patient in the SWEC-ETHZ and CHB-MIT datasets and its corresponding AUC ROC and AUC PR. For SWEC-ETHZ dataset, an unsupervised approach was the best-performing implementation for $7$ out of $12$ patients. %Table \ref{table:bestimpl_chb} shows the equivalent for the CHB-MIT dataset. 
For the CHB-MIT dataset, for $16$ out of $23$ patients, supervised approaches performed better. % as their best-performing implementation (16 out of 23). 
In particular, the supervised CNN performed the best for $8$ patients -- the most of any model. 
%Across both datasets, the top-performing model varies significantly depending on the patient. 21 patients had supervised approaches as their best implementation, while unsupervised approaches had 14. 
Figure \ref{fig:arch_breakdown} shows that using the CNN-LSTM was the most effective for the most patients, with the best performance in $16$ of the $35$ patients.

\begin{table*}[!htbp]
\centering
\begin{tabular}{ |c|c|c|c| } 
 \hline
 \textbf{Patient ID} & \textbf{Best Implementation} & \textbf{AUC PR} & \textbf{AUC ROC} \\ \hline
ID03    & Supervised CNN      & 0.735      & 0.930 \\ \hline
ID04    & Unsupervised CNN-LSTM        & 0.727      & 0.427 \\ \hline
ID05    & Unsupervised CNN-LSTM        & 0.991      & 0.995 \\ \hline
ID06    & Supervised CNN-LSTM & 0.537      & 0.756 \\ \hline
ID07    & Supervised CNN-LSTM & 0.726      & 0.787 \\ \hline
ID08    & Unsupervised TCN    & 0.501      & 0.778 \\ \hline
ID09    & Unsupervised CNN-LSTM        & 0.918      & 0.734 \\ \hline
ID10    & Unsupervised TCN    & 0.955      & 0.808 \\ \hline
ID12    & Unsupervised CNN    & 0.817      & 0.910 \\ \hline
ID13    & Supervised TCN      & 0.573      & 0.496 \\ \hline
ID16    & Unsupervised CNN    & 0.430      & 0.737 \\ \hline
ID18    & Supervised CNN      & 0.702      & 0.837 \\ \hline
\end{tabular}
\caption{Best-performing implementation for each patient in the SWEC-ETHZ dataset.}
\label{table:bestimpl_swec}
\end{table*}

\begin{table*}[!htbp]
\centering
\begin{tabular}{ |c|c|c|c| } 
 \hline
 \textbf{Patient ID} & \textbf{Best Implementation} & \textbf{AUC PR} & \textbf{ROC AUC} \\ \hline
 1   & Supervised CNN       & 0.966      & 0.988 \\ \hline
2   & Supervised CNN       & 0.737      & 0.968 \\ \hline
3   & Supervised CNN       & 1.000      & 1.000 \\ \hline
4   & Supervised TCN       & 0.064      & 0.751 \\ \hline
5   & Unsupervised CNN-LSTM & 0.612      & 0.891 \\ \hline
6   & Supervised CNN       & 0.878      & 0.925 \\ \hline
7   & Unsupervised CNN-LSTM & 0.115      & 0.526 \\ \hline
8   & Supervised TCN       & 0.704      & 0.773 \\ \hline
9   & Supervised CNN       & 0.660      & 0.947 \\ \hline
10  & Supervised CNN-LSTM  & 0.733      & 0.922 \\ \hline
11  & Supervised CNN       & 0.878      & 0.989 \\ \hline
12  & Unsupervised TCN      & 0.928      & 0.743 \\ \hline
13  & Supervised CNN       & 0.979      & 0.994 \\ \hline
14  & Supervised TCN       & 0.742      & 0.824 \\ \hline
15  & Unsupervised CNN-LSTM & 0.720      & 0.526 \\ \hline
16  & Unsupervised CNN-LSTM & 0.319      & 0.764 \\ \hline
17  & Unsupervised CNN-LSTM & 0.414      & 0.808 \\ \hline
18  & Supervised CNN-LSTM  & 0.104      & 0.822 \\ \hline
19  & Supervised CNN-LSTM  & 0.733      & 0.964 \\ \hline
20  & Supervised CNN-LSTM  & 0.706      & 0.924 \\ \hline
21  & Supervised CNN-LSTM  & 0.477      & 0.868 \\ \hline
22  & Unsupervised CNN-LSTM & 0.308      & 0.909 \\ \hline
23  & Supervised TCN       & 0.953      & 0.975 \\ \hline
\end{tabular}
\caption{Best-performing implementation for each patient in the CHB-MIT dataset.}
\label{table:bestimpl_chb}
\end{table*}

\begin{figure}[!htbp]
    \centering
    \includegraphics[scale=0.75]{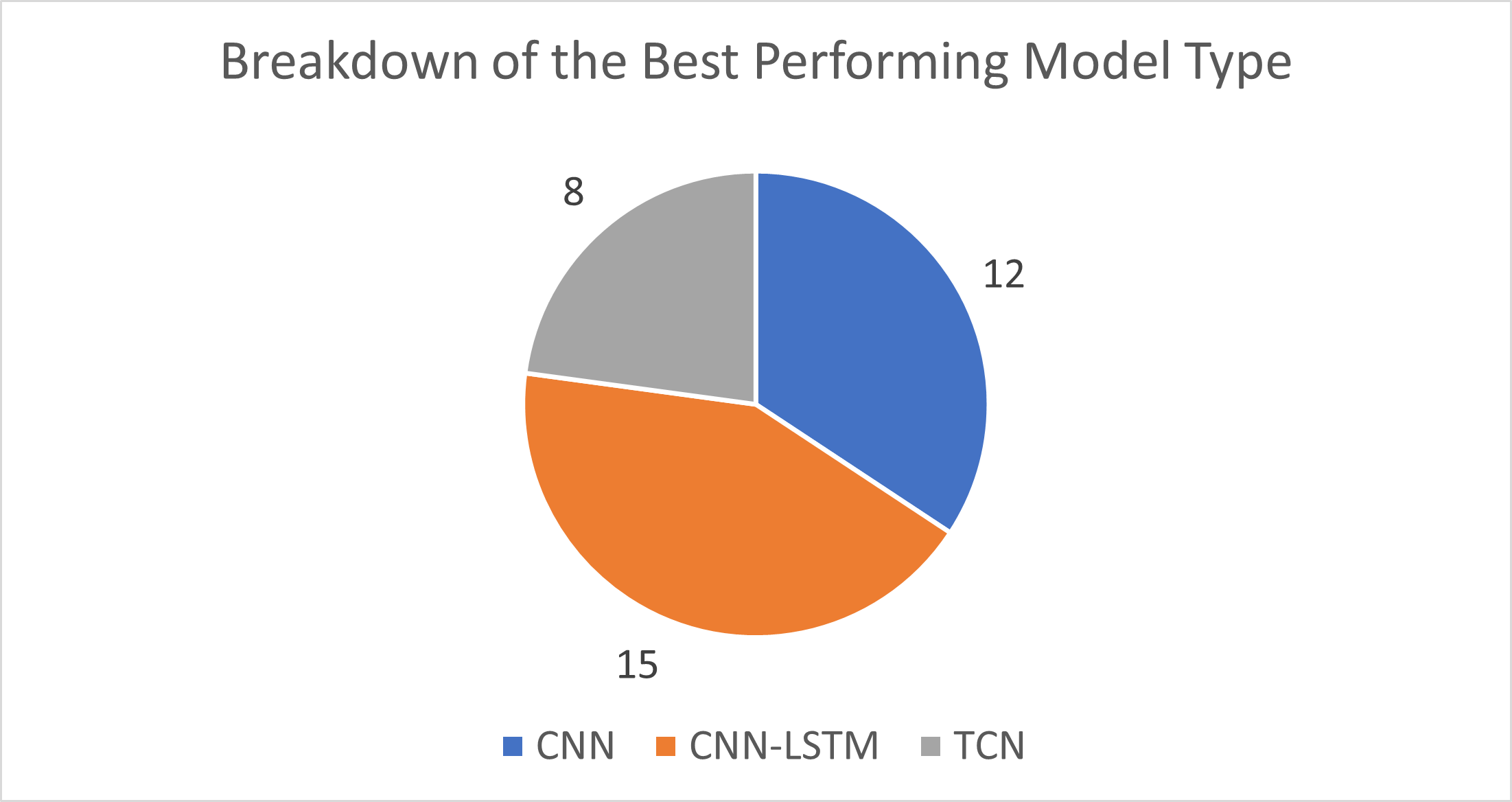} \\
    \caption{Number of times the best performing model was of each architecture type (CNN, CNN-LSTM, TCN).}
    \label{fig:arch_breakdown}
\end{figure}

\subsection{Discussion}
%There are a few important takeaways from this supervised seizure prediction work. 
We found that it is important to tune the window size and PPL to maximize performance. Preictal-interictal classification performed slightly better in both datasets when using an optimized hyperparameter implementation. However, in the CHB-MIT dataset, this difference was marginal. This is likely because of the size of the dataset. The CHB-MIT dataset has fewer data per patient compared to the SWEC-ETHZ dataset, so it is harder to properly tune hyperparameters that will generalize to new seizures.
%When comparing architectures, 
The CNN and CNN-LSTM architectures performed similarly in most experiments. This is likely because %their approach is similar. They 
both use time-frequency transforms followed by two-dimensional convolutions for spatial feature extraction. Even though the architectures are not exactly the same, it is likely that both are capturing similar underlying patterns in the data. The TCN performed fairly well and was able to get good results in some patients when the other two models failed. Although there was not one consistently high-performing model, it is encouraging that different architectures were able to perform well for different patients in different datasets.

The prediction results vary considerably across datasets, patients, and implementations. This demonstrates the variable nature of preictal and interictal data. To account for this, %it is important to have as much data as possible. In particular, 
it is important to have as many lead seizures data in a patient as possible, since preictal data is typically scarce. %It would be possible to train our models with a larger body of data, which may increase their ability to generalize to new seizures. It is also important so that we can have more data to test on. This would help account for a lot of the variability in the results. 
A limitation of our work is that a patient requires three lead seizures in their data to work with this method. It may not always be feasible for a patient's data to have at least three lead seizures, especially considering the difficulty of data acquisition.

Anomaly detection seizure prediction performance varied significantly across different architectures. Although supervised preictal-interictal classification performed better overall, there were many patients where an unsupervised approach was the best implementation. Additionally, in the SWEC-ETHZ dataset, an unsupervised approach was the best implementation for the majority of patients. This is likely because the SWEC-ETHZ dataset had a much larger recording duration and interictal-preictal ratio. %This is very encouraging for the methodology. 
In autoencoder based unsupervised approaches, the model is trained on only interictal (or normal) EEG data and reconstruction error is used to detect the onset of seizures (or preictal events). If the interictal data is interfered with noise, it means that the reconstruction error may be higher even for the interictal training data, which could result in misidentifying pre-seizures (that may also have higher reconstruction error because they were not seen before). Alternatively, if the interictal data is not diverse enough, then a slight variation in test interictal data could lead to misclassifying it as pre-seizure, which may lead to higher false alarms rate. Nevertheless, anomaly detection seizure prediction shows promise, and it implies that with improved signal processing and predictive modelling it may not be necessary to collect substantial preictal data to predict a seizure.

\begin{figure}[!htbp]
    \centering
    \includegraphics[scale=0.7]{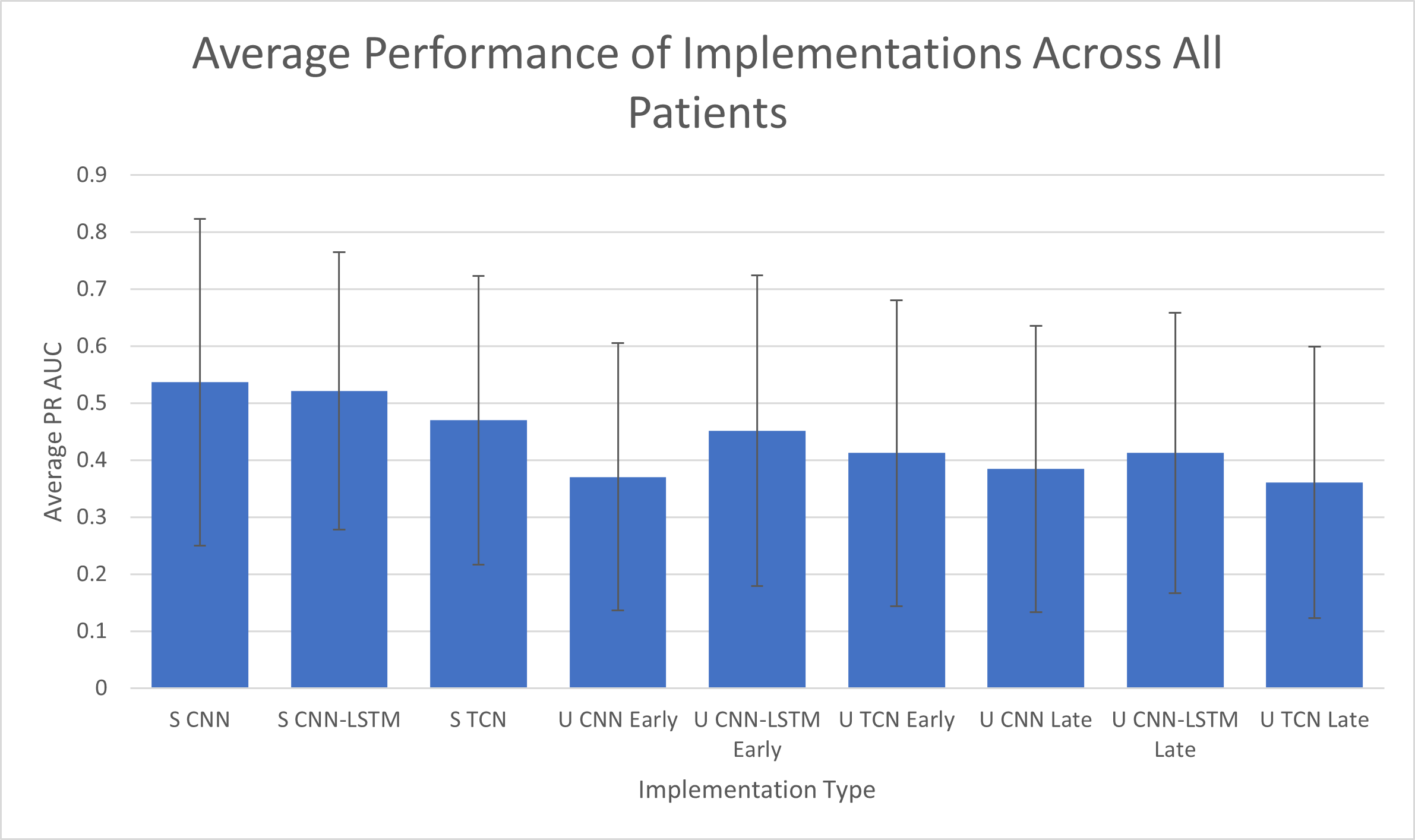} \\
    \caption{Average performance of each implementation. S: supervised. U: unsupervised.}
    \label{fig:avgperf}
\end{figure}
% \begin{figure}[h]
%     \centering
%     \includegraphics[scale=0.45]{Figures/medperf.png} \\
%     \caption{Median performance of each implementation. S: supervised. U: unsupervised. CNN: convolutional neural network. LSTM: long short-term neural network. TCN: temporal convolutional network.}
%     \label{fig:medperf}
% \end{figure}

Figure \ref{fig:avgperf} shows the average performance in terms of AUC PR across all patients. It can be observed that the supervised CNN and the supervised CNN-LSTM performed the best on average. However, the difference in performance across models is not large, and with a large standard deviation, it is impossible to make a statistical claim on the best performing model. %Figure \ref{fig:medperf} shows the median performance of each implementation across all patients. Again, the supervised CNN was the highest, followed by the supervised CNN-LSTM. 
In general, it can be observed that the supervised approaches performed better than the unsupervised approaches with % when looking at these graphs. 
%However, these 
results varying across individual patients.
Our results also showed the potential of using unsupervised approaches for seizure prediction. A major advantage is that it only uses unlabelled interictal EEG data, which is easier to acquire and is not dependent on an expert to annotate.

\section{Conclusions and Future Directions}
%Seizure prediction has remained a difficult problem to solve for decades. 
%Preictal-interictal classification has emerged as a popular method to predict the onset of a seizure. 
We developed several supervised approaches and introduced new unsupervised deep learning approaches for predicting epileptic seizures. In each approach, the main goal was to identify a preictal state (either as a binary class or anomaly) to predict the onset of an incoming seizure. We accounted for the variability of EEG and the preictal period by tuning the window size and PPL using a grid search. We trained personalized models and tuned hyperparameter using LOSO approach for better generalization of results. This method has achieved good results on more than half of the patients. We experimented with different supervised and unsupervised deep learning architectures on two large EEG datasets. Our results vary across different implementations depending on the patient.
The advantage of unsupervised methods is that they do not require preictal data to train the models; thus, alleviating the challenges around data acquisition, and effort and time spent in labelling. %Using different types of autoencoders, we trained our models solely on the interictal data and treated anything anomalous as an indicator of an impending seizure. 
However, due to the absence of validation set in unsupervised approaches (anomaly detection), it is non-trivial to obtain a threshold to detect pre-seizure events. Previous work on creating proxy outliers from the normal data can be extended to obtain an operating threshold for predicting seizures \cite{khan2017detecting, khan2023empirical}. We found that in many cases, an unsupervised approach was able to get similar or even better performance than a supervised approach; however, there was no single best performing model.
Our extensive experiments show the feasibility of supervised and unsupervised deep learning approaches for seizure prediction. However, the amount of preictal data per patient appears to be a crucial factor in training generalized models.

A future extension would be to experiment with a larger range for the hyperparameters. These parameters can also vary across implementations, so optimized hyperparameter implementations with the CNN-LSTM or TCN architecture as the base could be valuable. Another direction is to obtain (person-specific) operating threshold for deployment of these algorithms in a real-world setting. However, there are several limiting factors, including data imbalance, and unequal (and potentially unknown) costs of false positive and false negative in this application. These costs must be informed by clinical practices and guidelines as it pertains to human life. Generative adversarial networks \cite{habashi2023generative} and autoencoders trained in adversarial manner \cite{khan2021spatio} can be another potential unsupervised deep learning approach to predict epileptic seizures. Another extension would be to try different signal processing methods and advanced CNN and sequential models, including Resnet and Transformers. A breakthrough in reducing intervention time before the onset of seizure would lead to development of therapeutic interventions that can empower epilepsy patients to live without the fear or adversarial outcomes. 

\section*{Declarations}

\subsection*{Ethics Approval}
Two publicly available datasets were used in this work; therefore, ethics approvals were not required to use them.

\subsection*{Consent to Publish}
All the co-authors give their consent to publish this paper.

\subsection*{Competing interests}
The authors declare no competing interests of a financial or personal nature
 
\subsection*{Authors' contributions} 
Zakary Georgis-Yap and Shehroz S. Khan contributed equally in terms of developing algorithms, evaluation of results and manuscript preparation. Shehroz S. Khan prepared subsequent revisions of the paper. Milos R. Popovic provided academic support and consultation throughout the study.
 
\subsection*{Funding} 
This work is supported by the Natural Sciences and Engineering Research Council of Canada and Data Science Institute, University of Toronto. 
 
\subsection*{Availability of data and materials} 
The datasets used in the paper are publicly available.

\bibliography{references}


\begin{thebibliography}{55}
\ifx \bisbn   \undefined \def \bisbn  #1{ISBN #1}\fi
\ifx \binits  \undefined \def \binits#1{#1}\fi
\ifx \bauthor  \undefined \def \bauthor#1{#1}\fi
\ifx \batitle  \undefined \def \batitle#1{#1}\fi
\ifx \bjtitle  \undefined \def \bjtitle#1{#1}\fi
\ifx \bvolume  \undefined \def \bvolume#1{\textbf{#1}}\fi
\ifx \byear  \undefined \def \byear#1{#1}\fi
\ifx \bissue  \undefined \def \bissue#1{#1}\fi
\ifx \bfpage  \undefined \def \bfpage#1{#1}\fi
\ifx \blpage  \undefined \def \blpage #1{#1}\fi
\ifx \burl  \undefined \def \burl#1{\textsf{#1}}\fi
\ifx \doiurl  \undefined \def \doiurl#1{\url{https://doi.org/#1}}\fi
\ifx \betal  \undefined \def \betal{\textit{et al.}}\fi
\ifx \binstitute  \undefined \def \binstitute#1{#1}\fi
\ifx \binstitutionaled  \undefined \def \binstitutionaled#1{#1}\fi
\ifx \bctitle  \undefined \def \bctitle#1{#1}\fi
\ifx \beditor  \undefined \def \beditor#1{#1}\fi
\ifx \bpublisher  \undefined \def \bpublisher#1{#1}\fi
\ifx \bbtitle  \undefined \def \bbtitle#1{#1}\fi
\ifx \bedition  \undefined \def \bedition#1{#1}\fi
\ifx \bseriesno  \undefined \def \bseriesno#1{#1}\fi
\ifx \blocation  \undefined \def \blocation#1{#1}\fi
\ifx \bsertitle  \undefined \def \bsertitle#1{#1}\fi
\ifx \bsnm \undefined \def \bsnm#1{#1}\fi
\ifx \bsuffix \undefined \def \bsuffix#1{#1}\fi
\ifx \bparticle \undefined \def \bparticle#1{#1}\fi
\ifx \barticle \undefined \def \barticle#1{#1}\fi
\bibcommenthead
\ifx \bconfdate \undefined \def \bconfdate #1{#1}\fi
\ifx \botherref \undefined \def \botherref #1{#1}\fi
\ifx \url \undefined \def \url#1{\textsf{#1}}\fi
\ifx \bchapter \undefined \def \bchapter#1{#1}\fi
\ifx \bbook \undefined \def \bbook#1{#1}\fi
\ifx \bcomment \undefined \def \bcomment#1{#1}\fi
\ifx \oauthor \undefined \def \oauthor#1{#1}\fi
\ifx \citeauthoryear \undefined \def \citeauthoryear#1{#1}\fi
\ifx \endbibitem  \undefined \def \endbibitem {}\fi
\ifx \bconflocation  \undefined \def \bconflocation#1{#1}\fi
\ifx \arxivurl  \undefined \def \arxivurl#1{\textsf{#1}}\fi
\csname PreBibitemsHook\endcsname

%%% 1
\bibitem[\protect\citeauthoryear{Shahbazi and Aghajan}{2018}]{Moh18}
\begin{bchapter}
\bauthor{\bsnm{Shahbazi}, \binits{M.}},
\bauthor{\bsnm{Aghajan}, \binits{H.}}:
\bctitle{A generalizable model for seizure prediction based on deep learning
  using cnn-lstm architecture}.
In: \bbtitle{2018 IEEE Global Conference on Signal and Information Processing
  (GlobalSIP)},
\bconflocation{Anaheim, USA}
(\byear{2018})
\end{bchapter}
\endbibitem

%%% 2
\bibitem[\protect\citeauthoryear{Beghi and Giussani}{2018}]{Beg18}
\begin{barticle}
\bauthor{\bsnm{Beghi}, \binits{E.}},
\bauthor{\bsnm{Giussani}, \binits{G.}}:
\batitle{Aging and the epidemiology of epilepsy}.
\bjtitle{Neuroepidemiology}
\bvolume{51},
\bfpage{216}--\blpage{223}
(\byear{2018})
\end{barticle}
\endbibitem

%%% 3
\bibitem[\protect\citeauthoryear{Litt and Echauz}{2002}]{Lit02}
\begin{barticle}
\bauthor{\bsnm{Litt}, \binits{B.}},
\bauthor{\bsnm{Echauz}, \binits{J.}}:
\batitle{Prediction of epileptic seizures}.
\bjtitle{Neurology}
\bvolume{1},
\bfpage{22}--\blpage{30}
(\byear{2002})
\end{barticle}
\endbibitem

%%% 4
\bibitem[\protect\citeauthoryear{Nguyen and T'{e}llez~Zenteno}{2009}]{Ngu09}
\begin{botherref}
\oauthor{\bsnm{Nguyen}, \binits{R.}},
\oauthor{\bsnm{T'{e}llez~Zenteno}, \binits{J.F.}}:
Injuries in epilepsy: a review of its prevalence, risk factors, type of
  injuries and prevention.
Neurology International
\textbf{1}
(2009)
\end{botherref}
\endbibitem

%%% 5
\bibitem[\protect\citeauthoryear{Ridsdale et~al.}{2011}]{Rid11}
\begin{barticle}
\bauthor{\bsnm{Ridsdale}, \binits{L.}},
\bauthor{\bsnm{Charlton}, \binits{J.}},
\bauthor{\bsnm{Ashworth}, \binits{M.}},
\bauthor{\bsnm{Richardson}, \binits{M.P.}},
\bauthor{\bsnm{Gulliford}, \binits{M.C.}}:
\batitle{Epilepsy mortality and risk factors for death in epilepsy: a
  population-based study}.
\bjtitle{Br J Gen Pract.}
\bvolume{61},
\bfpage{271}--\blpage{278}
(\byear{2011})
\end{barticle}
\endbibitem

%%% 6
\bibitem[\protect\citeauthoryear{Fisher}{2000}]{Fis00}
\begin{barticle}
\bauthor{\bsnm{Fisher}, \binits{R.S.}}:
\batitle{Epilepsy from the patient's perspective: Review of results of a
  community-based survey}.
\bjtitle{Epilepsy \& Behavior}
\bvolume{1},
\bfpage{9}--\blpage{14}
(\byear{2000})
\end{barticle}
\endbibitem

%%% 7
\bibitem[\protect\citeauthoryear{Kuhlmann et~al.}{2018}]{Kuh18}
\begin{barticle}
\bauthor{\bsnm{Kuhlmann}, \binits{L.}},
\bauthor{\bsnm{Lehnertz}, \binits{K.}},
\bauthor{\bsnm{Richardson}, \binits{M.P.}},
\bauthor{\bsnm{Schelter}, \binits{B.}},
\bauthor{\bsnm{Zaveri}, \binits{H.P.}}:
\batitle{Seizure prediction --- ready for a new era}.
\bjtitle{Nature Reviews Neurology}
\bvolume{14},
\bfpage{618}--\blpage{630}
(\byear{2018})
\end{barticle}
\endbibitem

%%% 8
\bibitem[\protect\citeauthoryear{Stacey et~al.}{2011}]{Sta11}
\begin{barticle}
\bauthor{\bsnm{Stacey}, \binits{W.}},
\bauthor{\bsnm{Le~Van~Quyen}, \binits{M.}},
\bauthor{\bsnm{Mormann}, \binits{F.}},
\bauthor{\bsnm{Schulze-Bonhage}, \binits{A.}}:
\batitle{What is the present-day eeg evidence for a preictal state?}
\bjtitle{Epilepsy Research}
\bvolume{97},
\bfpage{243}--\blpage{251}
(\byear{2011})
\end{barticle}
\endbibitem

%%% 9
\bibitem[\protect\citeauthoryear{Brinkamann et~al.}{2016}]{Bri16}
\begin{barticle}
\bauthor{\bsnm{Brinkamann}, \binits{B.H.}},
\bauthor{\bsnm{Wagenaar}, \binits{J.}},
\bauthor{\bsnm{Abbot}, \binits{D.}},
\bauthor{\bsnm{Adkins}, \binits{P.}},
\bauthor{\bsnm{Bosshard}, \binits{S.C.}},
\bauthor{\bsnm{Chen}, \binits{M.}},
\bauthor{\bsnm{Tieng}, \binits{Q.M.}},
\bauthor{\bsnm{He}, \binits{J.}},
\bauthor{\bsnm{Mu\~{n}oz-Almaraz}, \binits{F.J.}},
\bauthor{\bsnm{Botella-Rocamora}, \binits{P.}},
\bauthor{\bsnm{Pardo}, \binits{J.}},
\bauthor{\bsnm{Zamora-Martinez}, \binits{F.}},
\bauthor{\bsnm{Hills}, \binits{M.}},
\bauthor{\bsnm{Wu}, \binits{W.}},
\bauthor{\bsnm{Korshunova}, \binits{I.}},
\bauthor{\bsnm{Cukierski}, \binits{W.}},
\bauthor{\bsnm{Vite}, \binits{C.}},
\bauthor{\bsnm{Patterson}, \binits{E.E.}},
\bauthor{\bsnm{Litt}, \binits{B.}},
\bauthor{\bsnm{Worrel}, \binits{G.A.}}:
\batitle{Crowdsourcing reproducible seizure forecasting in human and canine
  epilepsy}.
\bjtitle{Brain}
\bvolume{139},
\bfpage{1713}--\blpage{1722}
(\byear{2016})
\end{barticle}
\endbibitem

%%% 10
\bibitem[\protect\citeauthoryear{Bandarabadi et~al.}{2015}]{Ban15}
\begin{barticle}
\bauthor{\bsnm{Bandarabadi}, \binits{M.}},
\bauthor{\bsnm{Rasekhi}, \binits{J.}},
\bauthor{\bsnm{Teixeira}, \binits{C.A.}},
\bauthor{\bsnm{Karami}, \binits{M.R.}},
\bauthor{\bsnm{Dourado}, \binits{A.}}:
\batitle{On the proper selection of preictal period for seizure prediction}.
\bjtitle{Epilepsy \& Behavior}
\bvolume{158-166},
\bfpage{46}
(\byear{2015})
\end{barticle}
\endbibitem

%%% 11
\bibitem[\protect\citeauthoryear{Giannakakis et~al.}{2014}]{Gia14}
\begin{barticle}
\bauthor{\bsnm{Giannakakis}, \binits{G.}},
\bauthor{\bsnm{Sakkalis}, \binits{V.}},
\bauthor{\bsnm{Pediaditis}, \binits{M.}},
\bauthor{\bsnm{Tsiknakis}, \binits{M.}}:
\batitle{Methods for seizure detection and prediction: An overview}.
\bjtitle{Modern Electroencephalographic Assessment Techniques}
\bvolume{91},
\bfpage{131}--\blpage{157}
(\byear{2014})
\end{barticle}
\endbibitem

%%% 12
\bibitem[\protect\citeauthoryear{Georgis-Yap et~al.}{2022}]{georgispreictal}
\begin{bchapter}
\bauthor{\bsnm{Georgis-Yap}, \binits{Z.}},
\bauthor{\bsnm{Popovic}, \binits{M.R.}},
\bauthor{\bsnm{Khan}, \binits{S.S.}}:
\bctitle{Preictal-interictal classification for seizure prediction}.
In: \bbtitle{The 35th Canadian Conference on Artificial Intelligence}
(\byear{2022})
\end{bchapter}
\endbibitem

%%% 13
\bibitem[\protect\citeauthoryear{Burrello et~al.}{2019}]{Bur19}
\begin{bchapter}
\bauthor{\bsnm{Burrello}, \binits{A.}},
\bauthor{\bsnm{Cavigelli}, \binits{L.}},
\bauthor{\bsnm{Schindler}, \binits{K.}},
\bauthor{\bsnm{Benini}, \binits{L.}},
\bauthor{\bsnm{Rahimi}, \binits{A.}}:
\bctitle{Laelaps: An energy-efficient seizure detection algorithm from
  long-term human ieeg recordings without false alarms}.
In: \bbtitle{2019 Design, Automation \& Test in Europe Conference \& Exhibition
  (DATE)},
\bconflocation{Florence, Italy}
(\byear{2019})
\end{bchapter}
\endbibitem

%%% 14
\bibitem[\protect\citeauthoryear{Shoeb}{2009}]{Sho09}
\begin{botherref}
\oauthor{\bsnm{Shoeb}, \binits{A.H.}}:
Application of machine learning to epileptic seizure onset detection and
  treatment.
PhD thesis,
Massachusetts Institute of Technology
(2009)
\end{botherref}
\endbibitem

%%% 15
\bibitem[\protect\citeauthoryear{Acharya et~al.}{2018}]{Ach18}
\begin{barticle}
\bauthor{\bsnm{Acharya}, \binits{U.R.}},
\bauthor{\bsnm{Hagiwara}, \binits{Y.}},
\bauthor{\bsnm{Adeli}, \binits{H.}}:
\batitle{Automated seizure prediction}.
\bjtitle{Epilepsy \& Behavior}
\bvolume{88},
\bfpage{251}--\blpage{261}
(\byear{2018})
\end{barticle}
\endbibitem

%%% 16
\bibitem[\protect\citeauthoryear{Natu et~al.}{2022}]{natu2022review}
\begin{botherref}
\oauthor{\bsnm{Natu}, \binits{M.}},
\oauthor{\bsnm{Bachute}, \binits{M.}},
\oauthor{\bsnm{Gite}, \binits{S.}},
\oauthor{\bsnm{Kotecha}, \binits{K.}},
\oauthor{\bsnm{Vidyarthi}, \binits{A.}}:
Review on epileptic seizure prediction: machine learning and deep learning
  approaches.
Computational and Mathematical Methods in Medicine
\textbf{2022}
(2022)
\end{botherref}
\endbibitem

%%% 17
\bibitem[\protect\citeauthoryear{LeCun et~al.}{2015}]{LeC15}
\begin{barticle}
\bauthor{\bsnm{LeCun}, \binits{Y.}},
\bauthor{\bsnm{Bengio}, \binits{Y.}},
\bauthor{\bsnm{Hinton}, \binits{G.}}:
\batitle{Deep learning}.
\bjtitle{Nature}
\bvolume{521},
\bfpage{436}--\blpage{444}
(\byear{2015})
\end{barticle}
\endbibitem

%%% 18
\bibitem[\protect\citeauthoryear{Teixeira et~al.}{2014}]{Tei14}
\begin{barticle}
\bauthor{\bsnm{Teixeira}, \binits{C.A.}},
\bauthor{\bsnm{Direito}, \binits{B.}},
\bauthor{\bsnm{Bandarabadi}, \binits{M.}},
\bauthor{\bsnm{Le~Van~Quyen}, \binits{M.}},
\bauthor{\bsnm{Valerrama}, \binits{M.}},
\bauthor{\bsnm{Schelter}, \binits{B.}},
\bauthor{\bsnm{Schulze-Bonhage}, \binits{A.}},
\bauthor{\bsnm{Navarro}, \binits{V.}},
\bauthor{\bsnm{Sales}, \binits{F.}},
\bauthor{\bsnm{Dourado}, \binits{A.}}:
\batitle{Epileptic seizure predictors based on computational intelligence
  techniques: A comparative study with 278 patients}.
\bjtitle{Computer Methods and Programs in Biomedicine}
\bvolume{114},
\bfpage{324}--\blpage{336}
(\byear{2014})
\end{barticle}
\endbibitem

%%% 19
\bibitem[\protect\citeauthoryear{Fei et~al.}{2017}]{Fei17}
\begin{barticle}
\bauthor{\bsnm{Fei}, \binits{K.}},
\bauthor{\bsnm{Wang}, \binits{W.}},
\bauthor{\bsnm{Yang}, \binits{Q.}},
\bauthor{\bsnm{Tang}, \binits{S.}}:
\batitle{Chaos feature study in fractional fourier domain for preictal
  prediction of epileptic seizure}.
\bjtitle{Neurocomputing}
\bvolume{249},
\bfpage{290}--\blpage{298}
(\byear{2017})
\end{barticle}
\endbibitem

%%% 20
\bibitem[\protect\citeauthoryear{Mirowski et~al.}{2009}]{Mir09}
\begin{barticle}
\bauthor{\bsnm{Mirowski}, \binits{P.}},
\bauthor{\bsnm{Madhavan}, \binits{D.}},
\bauthor{\bsnm{LeCun}, \binits{Y.}},
\bauthor{\bsnm{Kuzniecky}, \binits{R.}}:
\batitle{Classification of patterns of eeg synchronization for seizure
  prediction}.
\bjtitle{Clinical Neurophysiology}
\bvolume{120},
\bfpage{1927}--\blpage{1940}
(\byear{2009})
\end{barticle}
\endbibitem

%%% 21
\bibitem[\protect\citeauthoryear{Khan et~al.}{2018}]{Kha18}
\begin{barticle}
\bauthor{\bsnm{Khan}, \binits{H.}},
\bauthor{\bsnm{Marcuse}, \binits{L.}},
\bauthor{\bsnm{Fields}, \binits{M.}},
\bauthor{\bsnm{Swann}, \binits{K.}},
\bauthor{\bsnm{Yener}, \binits{B.}}:
\batitle{Focal onset seizure prediction using convolutional networks}.
\bjtitle{IEEE Transactions on Biomedical Engineering}
\bvolume{65},
\bfpage{2109}--\blpage{2118}
(\byear{2018})
\end{barticle}
\endbibitem

%%% 22
\bibitem[\protect\citeauthoryear{Truong et~al.}{2018}]{Tru18}
\begin{barticle}
\bauthor{\bsnm{Truong}, \binits{N.D.}},
\bauthor{\bsnm{Nguyen}, \binits{A.D.}},
\bauthor{\bsnm{Kuhlmann}, \binits{L.}},
\bauthor{\bsnm{Bonyadi}, \binits{M.R.}},
\bauthor{\bsnm{Yang}, \binits{J.}},
\bauthor{\bsnm{Ippolito}, \binits{S.}},
\bauthor{\bsnm{Kavehei}, \binits{O.}}:
\batitle{Convolutional neural networks for seizure prediction using
  intracranial and scalp electroencephalogram}.
\bjtitle{Neural Networks}
\bvolume{105},
\bfpage{104}--\blpage{111}
(\byear{2018})
\end{barticle}
\endbibitem

%%% 23
\bibitem[\protect\citeauthoryear{Eberlein et~al.}{2018}]{Ebe18}
\begin{bchapter}
\bauthor{\bsnm{Eberlein}, \binits{M.}},
\bauthor{\bsnm{Hildebrand}, \binits{R.}},
\bauthor{\bsnm{Tetzlaff}, \binits{R.}},
\bauthor{\bsnm{Hoffmann}, \binits{N.}},
\bauthor{\bsnm{Kuhlmann}, \binits{L.}},
\bauthor{\bsnm{Brinkmann}, \binits{B.}},
\bauthor{\bsnm{M"{u}ller}, \binits{J.}}:
\bctitle{Convolutional neural networks for epileptic seizure prediction}.
In: \bbtitle{2018 IEEE International Conference on Bioinformatics and
  Biomedicine (BIBM)},
\bconflocation{Madrid}
(\byear{2018})
\end{bchapter}
\endbibitem

%%% 24
\bibitem[\protect\citeauthoryear{Zhang et~al.}{2020}]{Zha20}
\begin{barticle}
\bauthor{\bsnm{Zhang}, \binits{Y.}},
\bauthor{\bsnm{Guo}, \binits{Y.}},
\bauthor{\bsnm{Yang}, \binits{P.}},
\bauthor{\bsnm{Chen}, \binits{W.}},
\bauthor{\bsnm{Lo}, \binits{B.}}:
\batitle{Epilepsy seizure prediction on eeg using common spatial pattern and
  convolutional neural network}.
\bjtitle{IEEE Journal of Biomedical and Health Informatics}
\bvolume{24},
\bfpage{465}--\blpage{474}
(\byear{2020})
\end{barticle}
\endbibitem

%%% 25
\bibitem[\protect\citeauthoryear{Liu et~al.}{2020}]{YuL20}
\begin{bchapter}
\bauthor{\bsnm{Liu}, \binits{Y.}},
\bauthor{\bsnm{Sivathamboo}, \binits{S.}},
\bauthor{\bsnm{Goodin}, \binits{P.}},
\bauthor{\bsnm{Bonnington}, \binits{P.}},
\bauthor{\bsnm{Kwan}, \binits{P.}},
\bauthor{\bsnm{Kuhlmann}, \binits{L.}},
\bauthor{\bsnm{O'Brien}, \binits{T.}},
\bauthor{\bsnm{Perucca}, \binits{P.}},
\bauthor{\bsnm{Ge}, \binits{Z.}}:
\bctitle{Epileptic seizure detection using convolutional neural network: A
  multi-biosignal study}.
In: \bbtitle{ACSW '20: Proceedings of the Australasian Computer Science Week
  Multiconference},
\bconflocation{Melbourne, Australia}
(\byear{2020})
\end{bchapter}
\endbibitem

%%% 26
\bibitem[\protect\citeauthoryear{Dissanayake et~al.}{2021}]{Dis21}
\begin{barticle}
\bauthor{\bsnm{Dissanayake}, \binits{T.}},
\bauthor{\bsnm{Fernando}, \binits{T.}},
\bauthor{\bsnm{Denman}, \binits{S.}},
\bauthor{\bsnm{Sridharan}, \binits{S.}},
\bauthor{\bsnm{Fookes}, \binits{C.}}:
\batitle{Deep learning for patient-independent epileptic seizure prediction
  using scalp eeg signals}.
\bjtitle{IEEE Sensors Journal}
\bvolume{21},
\bfpage{9377}--\blpage{9388}
(\byear{2021})
\end{barticle}
\endbibitem

%%% 27
\bibitem[\protect\citeauthoryear{Jana and Mukherjee}{2021}]{Jan21}
\begin{botherref}
\oauthor{\bsnm{Jana}, \binits{R.}},
\oauthor{\bsnm{Mukherjee}, \binits{I.}}:
Deep learning based efficient epileptic seizure prediction with eeg channel
  optimization.
Biomedical Signal Processing and Control
\textbf{68}
(2021)
\end{botherref}
\endbibitem

%%% 28
\bibitem[\protect\citeauthoryear{Xu et~al.}{2020}]{XuY20}
\begin{bchapter}
\bauthor{\bsnm{Xu}, \binits{Y.}},
\bauthor{\bsnm{Yang}, \binits{J.}},
\bauthor{\bsnm{Zhao}, \binits{S.}},
\bauthor{\bsnm{Wu}, \binits{H.}},
\bauthor{\bsnm{Sawan}, \binits{M.}}:
\bctitle{An end-to-end deep learning approach for epileptic seizure
  prediction}.
In: \bbtitle{2020 2nd IEEE International Conference on Artificial Intelligence
  Circuits and Systems (AICAS)},
\bconflocation{Italy}
(\byear{2020})
\end{bchapter}
\endbibitem

%%% 29
\bibitem[\protect\citeauthoryear{Tsiouris et~al.}{2018}]{Tsi18}
\begin{barticle}
\bauthor{\bsnm{Tsiouris}, \binits{K.M.}},
\bauthor{\bsnm{Pezoulas}, \binits{V.C.}},
\bauthor{\bsnm{Zervakis}, \binits{M.}},
\bauthor{\bsnm{Konitsiotis}, \binits{S.}},
\bauthor{\bsnm{Koutsouris}, \binits{D.D.}},
\bauthor{\bsnm{Fotiadis}, \binits{D.I.}}:
\batitle{A long short-term memory deep learning network for the prediction of
  epileptic seizures using eeg signals}.
\bjtitle{Computers in Biology and Medicine}
\bvolume{99},
\bfpage{24}--\blpage{37}
(\byear{2018})
\end{barticle}
\endbibitem

%%% 30
\bibitem[\protect\citeauthoryear{Abdelhameed and Bayoumi}{2018}]{Abd18}
\begin{bchapter}
\bauthor{\bsnm{Abdelhameed}, \binits{A.M.}},
\bauthor{\bsnm{Bayoumi}, \binits{M.}}:
\bctitle{Semi-supervised deep learning system for epileptic seizures onset
  prediction}.
In: \bbtitle{2018 17th IEEE International Conference on Machine Learning and
  Applications (ICMLA)},
\bconflocation{Orlando}
(\byear{2018})
\end{bchapter}
\endbibitem

%%% 31
\bibitem[\protect\citeauthoryear{Daoud and Bayoumi}{2019}]{Dao19}
\begin{barticle}
\bauthor{\bsnm{Daoud}, \binits{H.}},
\bauthor{\bsnm{Bayoumi}, \binits{M.A.}}:
\batitle{Efficient epileptic seizure prediction based on deep learning}.
\bjtitle{IEEE Transactions on Biomedical Circuits and Systems}
\bvolume{13},
\bfpage{804}--\blpage{813}
(\byear{2019})
\end{barticle}
\endbibitem

%%% 32
\bibitem[\protect\citeauthoryear{Wei et~al.}{2019}]{Wei19}
\begin{botherref}
\oauthor{\bsnm{Wei}, \binits{X.}},
\oauthor{\bsnm{Zhou}, \binits{L.}},
\oauthor{\bsnm{Zhang}, \binits{Z.}},
\oauthor{\bsnm{Chen}, \binits{Z.}},
\oauthor{\bsnm{Zhou}, \binits{Y.}}:
Early prediction of epileptic seizures using a long-term recurrent
  convolutional network.
Journal of Neuroscience Methods
\textbf{327}
(2019)
\end{botherref}
\endbibitem

%%% 33
\bibitem[\protect\citeauthoryear{Usman et~al.}{2020}]{Usm20}
\begin{barticle}
\bauthor{\bsnm{Usman}, \binits{S.M.}},
\bauthor{\bsnm{Khalid}, \binits{S.}},
\bauthor{\bsnm{Aslam}, \binits{M.H.}}:
\batitle{Epileptic seizures prediction using deep learning techniques}.
\bjtitle{IEEE Access}
\bvolume{8},
\bfpage{39998}--\blpage{40007}
(\byear{2020})
\end{barticle}
\endbibitem

%%% 34
\bibitem[\protect\citeauthoryear{Hussein et~al.}{2020}]{Hus20}
\begin{botherref}
\oauthor{\bsnm{Hussein}, \binits{A.}},
\oauthor{\bsnm{Djandji}, \binits{M.}},
\oauthor{\bsnm{Mahmoud}, \binits{R.}},
\oauthor{\bsnm{Dhaybi}, \binits{M.}},
\oauthor{\bsnm{Hajj}, \binits{H.M.}}:
Augmenting dl with adversarial training for robust prediction of epilepsy
  seizures.
Journal of the ACM
\textbf{1}
(2020)
\end{botherref}
\endbibitem

%%% 35
\bibitem[\protect\citeauthoryear{Prathaban and Balasubramanian}{2021}]{Pra21}
\begin{botherref}
\oauthor{\bsnm{Prathaban}, \binits{B.P.}},
\oauthor{\bsnm{Balasubramanian}, \binits{R.}}:
Dynamic learning framework for epileptic seizure prediction using sparsity
  based eeg reconstruction with optimized cnn classifier.
Expert Systems with Applications
\textbf{170}
(2021)
\end{botherref}
\endbibitem

%%% 36
\bibitem[\protect\citeauthoryear{Ozcan and Erturk}{2019}]{Ozc19}
\begin{barticle}
\bauthor{\bsnm{Ozcan}, \binits{A.R.}},
\bauthor{\bsnm{Erturk}, \binits{S.}}:
\batitle{Seizure prediction in scalp eeg using 3d convolutional neural networks
  with an image-based approach}.
\bjtitle{IEEE Transactions on Neural Systems and Rehabilitation Engineering}
\bvolume{27},
\bfpage{2284}--\blpage{2293}
(\byear{2019})
\end{barticle}
\endbibitem

%%% 37
\bibitem[\protect\citeauthoryear{Truong et~al.}{2019}]{Tru19}
\begin{bchapter}
\bauthor{\bsnm{Truong}, \binits{N.D.}},
\bauthor{\bsnm{Zhou}, \binits{L.}},
\bauthor{\bsnm{Kavehei}, \binits{O.}}:
\bctitle{Semi-supervised seizure prediction with generative adversarial
  networks}.
In: \bbtitle{2019 41st Annual International Conference of the IEEE Engineering
  in Medicine and Biology Society (EMBC)},
\bconflocation{Berlin}
(\byear{2019})
\end{bchapter}
\endbibitem

%%% 38
\bibitem[\protect\citeauthoryear{Sherstinsky}{2020}]{sherstinsky2020fundamentals}
\begin{barticle}
\bauthor{\bsnm{Sherstinsky}, \binits{A.}}:
\batitle{Fundamentals of recurrent neural network (rnn) and long short-term
  memory (lstm) network}.
\bjtitle{Physica D: Nonlinear Phenomena}
\bvolume{404},
\bfpage{132306}
(\byear{2020})
\end{barticle}
\endbibitem

%%% 39
\bibitem[\protect\citeauthoryear{Thill et~al.}{2021}]{Thi21}
\begin{barticle}
\bauthor{\bsnm{Thill}, \binits{M.}},
\bauthor{\bsnm{Konen}, \binits{W.}},
\bauthor{\bsnm{Wang}, \binits{H.}},
\bauthor{\bsnm{B"{a}ck}, \binits{T.}}:
\batitle{Temporal convolutional autoencoder for unsupervised anomaly detection
  in time series}.
\bjtitle{Applied Soft Computing}
\bvolume{112},
\bfpage{107751}
(\byear{2021})
\end{barticle}
\endbibitem

%%% 40
\bibitem[\protect\citeauthoryear{Stefan~Denkovski}{2023}]{Denkovski2023}
\begin{bchapter}
\bauthor{\bsnm{Stefan~Denkovski}, \binits{S.S.K.} \bsuffix{Alex~Mihailidis}}:
\bctitle{Temporal shift - multi-objective loss function for improved anomaly
  fall detection}.
In: \bbtitle{$15^{th}$ Asian Conference on Machine Learning}
(\byear{2023})
\end{bchapter}
\endbibitem

%%% 41
\bibitem[\protect\citeauthoryear{Khan et~al.}{2021}]{khan2021anomaly}
\begin{bchapter}
\bauthor{\bsnm{Khan}, \binits{S.S.}},
\bauthor{\bsnm{Khoshbakhtian}, \binits{F.}},
\bauthor{\bsnm{Ashraf}, \binits{A.B.}}:
\bctitle{Anomaly detection approach to identify early cases in a pandemic using
  chest x-rays.}
In: \bbtitle{Canadian Conference on AI}
(\byear{2021})
\end{bchapter}
\endbibitem

%%% 42
\bibitem[\protect\citeauthoryear{Jacob~Nogas and
  Mihailidis}{2018}]{nogas2018fall}
\begin{bchapter}
\bauthor{\bsnm{Jacob~Nogas}, \binits{S.S.K.}},
\bauthor{\bsnm{Mihailidis}, \binits{A.}}:
\bctitle{Fall detection from thermal camera using convolutional lstm
  autoencoder}.
In: \bbtitle{Proceedings of the 2nd Workshop on Aging, Rehabilitation and
  Independent Assisted Living, IJCAI Workshop}
(\byear{2018})
\end{bchapter}
\endbibitem

%%% 43
\bibitem[\protect\citeauthoryear{Abedi and Khan}{2023}]{abedi2022detecting}
\begin{botherref}
\oauthor{\bsnm{Abedi}, \binits{A.}},
\oauthor{\bsnm{Khan}, \binits{S.S.}}:
Detecting disengagement in virtual learning as an anomaly using temporal
  convolutional network autoencoder.
Signal, Image and Video Processing
(2023)
\end{botherref}
\endbibitem

%%% 44
\bibitem[\protect\citeauthoryear{Al-Fahoum and Al-Fraihat}{2014}]{AlF14}
\begin{botherref}
\oauthor{\bsnm{Al-Fahoum}, \binits{A.S.}},
\oauthor{\bsnm{Al-Fraihat}, \binits{A.A.}}:
Methods of eeg signal features extraction using linear analysis in frequency
  and time-frequency domains.
International Scholarly Research Notices
\textbf{2014}
(2014)
\end{botherref}
\endbibitem

%%% 45
\bibitem[\protect\citeauthoryear{Herrmann et~al.}{2014a}]{Her14}
\begin{barticle}
\bauthor{\bsnm{Herrmann}, \binits{C.S.}},
\bauthor{\bsnm{Rach}, \binits{S.}},
\bauthor{\bsnm{Vosskuhl}, \binits{J.}},
\bauthor{\bsnm{Str"{u}ber}, \binits{D.}}:
\batitle{Time--frequency analysis of event-related potentials: A brief
  tutorial}.
\bjtitle{Brain Topography}
\bvolume{27},
\bfpage{438}--\blpage{450}
(\byear{2014})
\end{barticle}
\endbibitem

%%% 46
\bibitem[\protect\citeauthoryear{Herrmann et~al.}{2014b}]{Her141}
\begin{barticle}
\bauthor{\bsnm{Herrmann}, \binits{C.S.}},
\bauthor{\bsnm{Rach}, \binits{S.}},
\bauthor{\bsnm{Vosskuhl}, \binits{J.}},
\bauthor{\bsnm{Str"{u}ber}, \binits{D.}}:
\batitle{Time--frequency analysis of event-related potentials: A brief
  tutorial}.
\bjtitle{Brain Topography}
\bvolume{27},
\bfpage{438}--\blpage{450}
(\byear{2014})
\end{barticle}
\endbibitem

%%% 47
\bibitem[\protect\citeauthoryear{Fadzal et~al.}{2012}]{Fad12}
\begin{bchapter}
\bauthor{\bsnm{Fadzal}, \binits{C.W.N.F.C.W.}},
\bauthor{\bsnm{Mansor}, \binits{W.}},
\bauthor{\bsnm{Khuan}, \binits{L.Y.}},
\bauthor{\bsnm{Zabidi}, \binits{A.}}:
\bctitle{Short-time fourier transform analysis of eeg signal from writing}.
In: \bbtitle{2012 IEEE 8th International Colloquium on Signal Processing and
  Its Applications},
\bconflocation{Malacca, Malaysia}
(\byear{2012})
\end{bchapter}
\endbibitem

%%% 48
\bibitem[\protect\citeauthoryear{Edakawa et~al.}{2016}]{edakawa2016detection}
\begin{barticle}
\bauthor{\bsnm{Edakawa}, \binits{K.}},
\bauthor{\bsnm{Yanagisawa}, \binits{T.}},
\bauthor{\bsnm{Kishima}, \binits{H.}},
\bauthor{\bsnm{Fukuma}, \binits{R.}},
\bauthor{\bsnm{Oshino}, \binits{S.}},
\bauthor{\bsnm{Khoo}, \binits{H.M.}},
\bauthor{\bsnm{Kobayashi}, \binits{M.}},
\bauthor{\bsnm{Tanaka}, \binits{M.}},
\bauthor{\bsnm{Yoshimine}, \binits{T.}}:
\batitle{Detection of epileptic seizures using phase--amplitude coupling in
  intracranial electroencephalography}.
\bjtitle{Scientific reports}
\bvolume{6}(\bissue{1}),
\bfpage{1}--\blpage{8}
(\byear{2016})
\end{barticle}
\endbibitem

%%% 49
\bibitem[\protect\citeauthoryear{Huang and Ling}{2005}]{Hua05}
\begin{barticle}
\bauthor{\bsnm{Huang}, \binits{J.}},
\bauthor{\bsnm{Ling}, \binits{C.X.}}:
\batitle{Using auc and accuracy in evaluating learning algorithms}.
\bjtitle{IEEE Transactions on Knowledge and Data Engineering}
\bvolume{17},
\bfpage{299}--\blpage{310}
(\byear{2005})
\end{barticle}
\endbibitem

%%% 50
\bibitem[\protect\citeauthoryear{Paszke et~al.}{2019}]{Pas19}
\begin{bchapter}
\bauthor{\bsnm{Paszke}, \binits{A.}},
\bauthor{\bsnm{Gross}, \binits{S.}},
\bauthor{\bsnm{Massa}, \binits{F.}},
\bauthor{\bsnm{Lerer}, \binits{A.}},
\bauthor{\bsnm{Bradbury}, \binits{J.}},
\bauthor{\bsnm{Chanan}, \binits{G.}},
\bauthor{\bsnm{Killeen}, \binits{T.}},
\bauthor{\bsnm{Lin}, \binits{Z.}},
\bauthor{\bsnm{Gimelshein}, \binits{N.}},
\bauthor{\bsnm{Antiga}, \binits{L.}},
\bauthor{\bsnm{Desmaison}, \binits{A.}},
\bauthor{\bsnm{Kopf}, \binits{A.}},
\bauthor{\bsnm{Yang}, \binits{E.}},
\bauthor{\bsnm{DeVito}, \binits{Z.}},
\bauthor{\bsnm{Raison}, \binits{M.}},
\bauthor{\bsnm{Tajani}, \binits{A.}},
\bauthor{\bsnm{Chilamkurthy}, \binits{S.}},
\bauthor{\bsnm{Steiner}, \binits{B.}},
\bauthor{\bsnm{Fang}, \binits{L.}},
\bauthor{\bsnm{Bai}, \binits{J.}},
\bauthor{\bsnm{Chintala}, \binits{S.}}:
\bctitle{Pytorch: An imperative style, high-performance deep learning library}.
In: \bbtitle{Advances in Neural Information Processing Systems 32},
\bconflocation{Vancouver}
(\byear{2019})
\end{bchapter}
\endbibitem

%%% 51
\bibitem[\protect\citeauthoryear{Branco et~al.}{2016}]{Bra16}
\begin{barticle}
\bauthor{\bsnm{Branco}, \binits{P.}},
\bauthor{\bsnm{Torgo}, \binits{L.}},
\bauthor{\bsnm{Ribeiro}, \binits{R.R.}}:
\batitle{A survey of predictive modeling on imbalanced domains}.
\bjtitle{ACM Computing Surveys (CSUR)}
\bvolume{49},
\bfpage{1}--\blpage{50}
(\byear{2016})
\end{barticle}
\endbibitem

%%% 52
\bibitem[\protect\citeauthoryear{Khan et~al.}{2017}]{khan2017detecting}
\begin{barticle}
\bauthor{\bsnm{Khan}, \binits{S.S.}},
\bauthor{\bsnm{Karg}, \binits{M.E.}},
\bauthor{\bsnm{Kuli{\'c}}, \binits{D.}},
\bauthor{\bsnm{Hoey}, \binits{J.}}:
\batitle{Detecting falls with x-factor hidden markov models}.
\bjtitle{Applied Soft Computing}
\bvolume{55},
\bfpage{168}--\blpage{177}
(\byear{2017})
\end{barticle}
\endbibitem

%%% 53
\bibitem[\protect\citeauthoryear{Khan et~al.}{2023}]{khan2023empirical}
\begin{bchapter}
\bauthor{\bsnm{Khan}, \binits{S.S.}},
\bauthor{\bsnm{Mishra}, \binits{P.K.}},
\bauthor{\bsnm{Ye}, \binits{B.}},
\bauthor{\bsnm{Newman}, \binits{K.}},
\bauthor{\bsnm{Iaboni}, \binits{A.}},
\bauthor{\bsnm{Mihailidis}, \binits{A.}}:
\bctitle{Empirical thresholding on spatio-temporal autoencoders trained on
  surveillance videos in a dementia care unit}.
In: \bbtitle{2023 20th Conference on Robots and Vision (CRV)},
pp. \bfpage{265}--\blpage{272}
(\byear{2023}).
\bcomment{IEEE}
\end{bchapter}
\endbibitem

%%% 54
\bibitem[\protect\citeauthoryear{Habashi et~al.}{2023}]{habashi2023generative}
\begin{barticle}
\bauthor{\bsnm{Habashi}, \binits{A.G.}},
\bauthor{\bsnm{Azab}, \binits{A.M.}},
\bauthor{\bsnm{Eldawlatly}, \binits{S.}},
\bauthor{\bsnm{Aly}, \binits{G.M.}}:
\batitle{Generative adversarial networks in eeg analysis: an overview}.
\bjtitle{Journal of NeuroEngineering and Rehabilitation}
\bvolume{20}(\bissue{1}),
\bfpage{40}
(\byear{2023})
\end{barticle}
\endbibitem

%%% 55
\bibitem[\protect\citeauthoryear{Khan et~al.}{2021}]{khan2021spatio}
\begin{barticle}
\bauthor{\bsnm{Khan}, \binits{S.S.}},
\bauthor{\bsnm{Nogas}, \binits{J.}},
\bauthor{\bsnm{Mihailidis}, \binits{A.}}:
\batitle{Spatio-temporal adversarial learning for detecting unseen falls}.
\bjtitle{Pattern Analysis and Applications}
\bvolume{24},
\bfpage{381}--\blpage{391}
(\byear{2021})
\end{barticle}
\endbibitem

\end{thebibliography}

\end{document}